\documentclass[aps,pra,twocolumn, notitlepage, superscriptaddress,nofootinbib]{revtex4-1}

\usepackage{color}
\usepackage{graphicx}
\usepackage{amsmath}
\usepackage{braket}
\usepackage{amssymb}
\usepackage{amsthm}

\usepackage{dsfont}

\newcommand{\haf}[0]{\mathrm{Haf}}

\newcommand{\upb}{Integrated Quantum Optics, Universit\"at Paderborn, Warburger Strasse 100, 33098 Paderborn, Germany}
\newcommand{\prague}{FNSPE, Czech Technical University in Prague, Br\^ehov\'{a} 7, 119 15, Praha 1, Czech Republic}

\begin{document}
\title{A detailed study of Gaussian Boson Sampling}

\author{Regina Kruse} \email{regina.kruse@upb.de}
\affiliation{\upb}
\author{Craig S. Hamilton} \email{hamilcra@fjfi.cvut.cz}
\affiliation{%
\prague
}%
\author{Linda Sansoni}
\affiliation{\upb}
\author{Sonja Barkhofen}
\affiliation{\upb}
\author{Christine Silberhorn}
\affiliation{\upb}
\author{Igor Jex }
\affiliation{%
\prague
}%

\begin{abstract}
Since the development of Boson sampling, there has been a quest to construct more efficient and experimentally feasible protocols to test the computational complexity of sampling from photonic states. In this paper we interpret and extend the results presented in [\textit{Phys. Rev. Lett. 119, 170501 (2017)}]. We derive an expression that relates the probability to measure a specific photon output pattern from a Gaussian state to the \textit{hafnian} matrix function and us it to design a Gaussian Boson sampling protocol. Then, we discuss the advantages that this protocol has relative to other photonic protocols and the experimental requirements for Gaussian Boson Sampling. Finally, we relate it to the previously most general protocol, Scattershot Boson Sampling [\textit{Phys. Rev. Lett. 113, 100502 (2014)}]. 
\end{abstract}

\maketitle

\section{Introduction}

Boson Sampling, introduced by Aaronson and Arkhipov (AABS), \cite{aaronson_computational_2011, Aaronson:2013p7598} is a non-universal model of quantum computation that may, for the first time, show the advantage of quantum-computational schemes over classical algorithms. From a computational point of view it is especially interesting, as it may provide evidence against the extended Church-Turing Thesis and experimentally it is attractive as it requires a straightforward implementation; $N$ single photon Fock states are launched in an $N^2$-dimensional linear interferometer and the output pattern of photons is measured. This experimental feasibility has inspired many groups to implement proof-of-principle experiments to demonstrate the viability of this protocol \cite{Broome:2013p7136, Tillmann:2013p10461, Spring:2013p7137, cres13npo2}. However, due to a lack of deterministic single-photon sources, these implementations had to use probabilistic, post-selected photon pair sources (Post-selected Fock Boson Sampling PFBS). The use of probabilistic sources means that the probability to generate high photon numbers in these schemes scales exponentially badly. Since a Boson Sampling experiment that may provide evidence against the extended Church-Turing thesis requires between $N=50-100$ photons \cite{aaronson_computational_2011, neville2017classical, clifford2017}, a probabilistic approach to photon generation is not likely to reach this benchmark. 

To improve the performance of the Boson sampling machines, groups have on one hand concentrated on the development of on-demand single photon sources to overcome the probabilistic nature of photon generation \cite{wang2017high, he_scalable_2016, loredo_boson_2017}. On the other hand, alternative and more feasible protocols were proposed \cite{Lund:2014p10967, Barkhofen:2017p13761}.
Scattershot Boson Sampling (SBS), proposed by Lund et al \cite{Lund:2014p10967} is a way to avoid the exponential scaling of probabilistic sources. This protocol makes use of $N^2$ two-mode squeezed states to generate $N$ photon pairs, where one photon of each pair acts as a herald for the other photon, which enters the input of the interferometer. These latter photons are the ones that are `sampled' in the AABS protocol. This increase in the number of resources improves the generation probability to a polynomial scaling for large photon numbers. 
An alternative method of utilising more sources was shown in \cite{Barkhofen:2017p13761}, yielding further improvements in the generation probability of photons. 

Although SBS and PFBS use weakly-squeezed Gaussian states (mean number of photons $\langle n \rangle \ll 1$) as the photon generation resource, these approaches reduce the protocol to sampling from single photon Fock states and do not exploit the full Gaussian nature of their initial states. 
This posed the question, from both theoretical and experimental perspectives, if a hybrid approach considering on the full gaussian nature of the input states and photon counting measurement schemes can improve existing sampling protocols. Such an approach benefits from the methods and concepts developed in the framework of both continuous and discrete variables quantum information, as Gaussian states are the basis of continuous variable quantum information and have been demonstrated to be a powerful resource for highly scalable systems, for example in the context of cluster state generation \cite{yoshikawa2016}. The special case of sampling from thermal states was answered in \cite{RahimiKeshari:2015p11006} and shown to be in $BPP^{NP}$, whereas measuring photons from coherent states is well-known and is in the simplest complexity class, P.

In a recent paper \cite{hamilton_gaussian_2016} we introduced Gaussian Boson Sampling (GBS) that answers questions about the complexity of sampling from a general squeezed state. There we derived a new expression that connects the probability to measure a specific output pattern of photons from a general Gaussian state to the hafnian matrix function. This was then used to develop a new regime of Boson sampling from squeezed states, which has specific advantages when compared to previous regimes i.e. SBS. In this paper, we extend our formula to account for displaced squeezed states (the addition of coherent light) and higher-order photon number contributions in a single output mode. Next, we go into detail on the construction of our GBS protocol with single mode squeezed states, discuss why the computation of the hafnian is in the \textsf{\#P} complexity class and provide arguments, similar to AA \cite{aaronson_computational_2011}, that approximate GBS is still in \textsf{\#P}. From this discussion, we derive several requirements on the experimental parameters and finally relate our GBS protocol to the most efficiently known Boson sampling protocol SBS. We then show that SBS is a special subclass of GBS protocols and demonstrate that GBS provides significant experimental advantages over current experimental realisations.

Our paper is structured as follows. In Sec. \ref{sec:review}, we review the main points of the AABS protocol. In Sec. \ref{sec:hafnian}, we derive the closed-formula expression that connects the probability to measure a specific photon output pattern from a general Gaussian state with the hafnian of a submatrix that is related to the covariance matrix of that state. Next, in Sec. \ref{sec:GBS}, we comment on the complexity of the hafnian and go into detail on the construction of our GBS protocol with single-mode squeezed states. Sec. \ref{sec:approx} summarises our arguments for the hardness of approximate GBS and we derive several requirements for an experiment in Sec. \ref{sec:requirements}. Additionally, in Sec. \ref{sec:GBS-vs-SBS}, we show that the most general protocol up to date SBS, is a specialised subclass of GBS problems and compare the experimental feasibility of our GBS protocol with existing experimental approaches in Sec. \ref{sec:efficiency}. Finally, we give some conclusions in Sec. \ref{sec:conclusion}.


%
\section{Review: AABS}\label{sec:review}
In this section we briefly review the original proposal by Aaronson and Arkhipov AABS \cite{aaronson_computational_2011, Aaronson:2013p7598}. Specifically, we are interested in an outline of their hardness proof, as  we base our arguments for approximate GBS (Sec. \ref{sec:approx}) on this.
\begin{figure}
\includegraphics[width=1.\columnwidth]{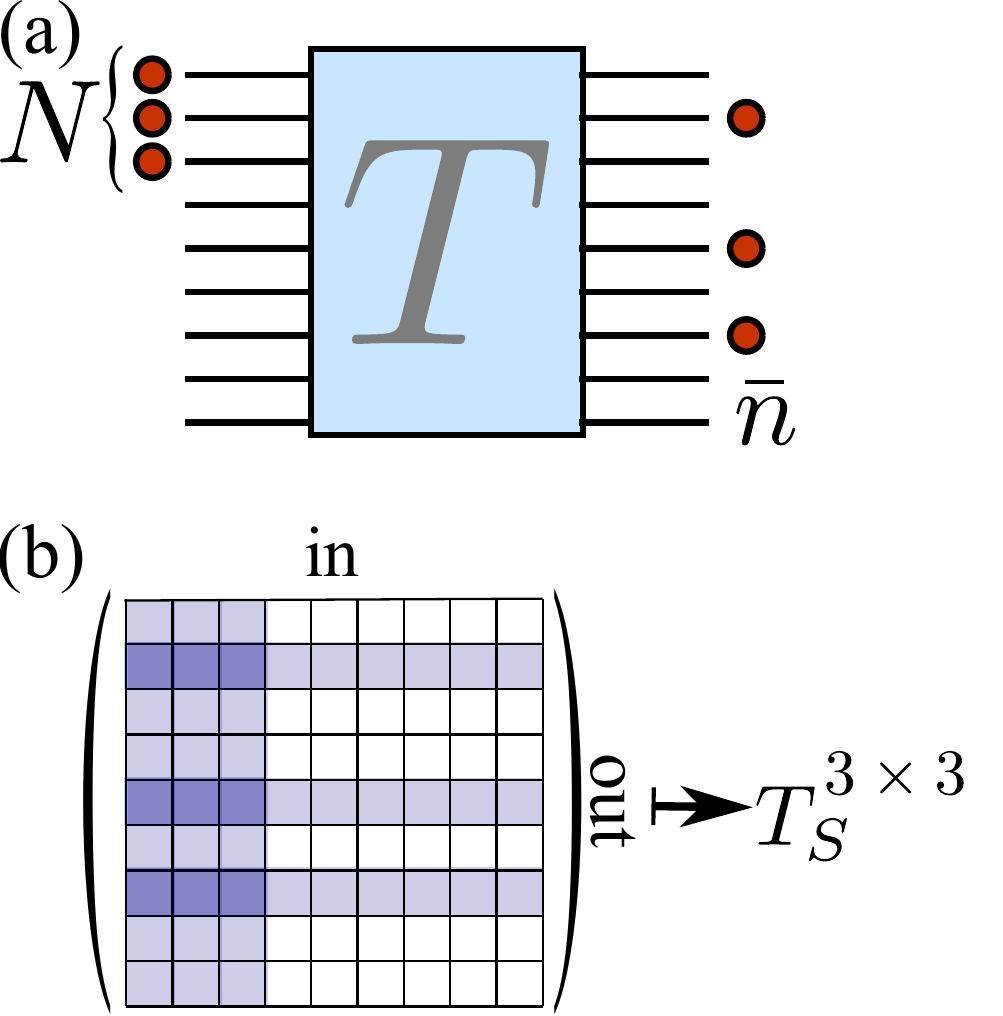}
\caption{(a) AABS scheme: $N$ photons are injected in the first $N$ input modes of an interferometer $\mathbf T$ and the output patterns $\bar{n}$ are sampled. The corresponding probability for a particular pattern $\bar{n}$ depends on the permanent of the sampled submatrix $\mathbf T_S$. (b) A typical construction of the sampled submatrix $\mathbf T_S$ for three photons where the first $3$ columns are preset by the input modes and the rows are selected by the output pattern $\bar{n}$. The matrix elements given by their intersections define the submatrix $\mathbf T_S$.}
\label{fig:AABS}
\end{figure}
For the AABS scheme, shown in Fig. \ref{fig:AABS}(a), $N$ pure, single photons are inserted into the first $N$ modes of an $M=\mathcal{O}(N^2)$-dimensional Haar random interferometer  $\mathbf T$.
At the output, we measure the number of photons in each mode, thereby sampling the output probability distribution of the device. It is assumed that all the photons leave in different modes, giving $M \choose N$ different output patterns. 
The probability to measure a specific pattern $\bar{n}$ is given by the permanent of the sampled submatrix of $\mathbf T$, which we call $\mathbf T_S$
\begin{equation}
\mathrm{Pr}(\bar{n})=|\mathrm{Perm}(T_S)|^2= \left|\sum_{\sigma\in P_N} \prod_{i=1}^N T_{S_{i,\sigma(i)}}\right|^2\, .
\label{eq:perm-probability}
\end{equation}
Here, $P_N$ are all permutations of size $N$. 
The process for constructing the submatrix $\mathbf T_S$ is illustrated for three photons in Fig. \ref{fig:AABS}(b). We select the columns of $\mathbf T$ corresponding to the position of the input photons and the rows of $\mathbf T$ corresponding to the output positions \cite{scheel2004permanents}. It is the intersection of these rows and columns that selects the entries of the matrix $\mathbf T_S$.

The main idea behind AABS is that the permanent is, in computational complexity theory, a \textsf{\#P}-complete problem, which means that it cannot be efficiently computed on a classical machine. Therefore, the calculation of all output pattern probabilities should also fall into the \textsf{\#P} complexity class and thus the output of the device cannot be efficiently sampled by a classical machine. To prove this claim, AA prove two main theorems, one for the exact sampling from such a distribution (i.e. from the exact probability distribution $\mathcal{D}_A$) and one for approximate sampling (from an approximation of $\mathcal{D}_A$, i.e. $\mathcal{D}'_A$). The proof for the first theorem mostly relies on the proof that the approximation of $\mathrm{Perm}(X)^2$ of a chosen matrix $X$ up to a multiplicative constant is a \textsf{\#P}-complete problem. In Sec. \ref{sec:approx}, we recall the main arguments of their complexity proof for the second theorem, i.e. the approximate sampling, and introduce arguments for approximate GBS, one which is based upon the AA proof and another that is unique to GBS.


%
\section{Photo-counts from a Gaussian state}\label{sec:hafnian}
In this section we consider the probability to measure photo-counts from a general Gaussian state, and derive the closed-formula expression for the probability to measure a specific photon output pattern $\bar{n}$. We showed in \cite{hamilton_gaussian_2016} that this probability is related to the hafnian \cite{Caianiello:1953p12510, Caianiello73} of a submatrix $\mathbf{A}_S$, which combines the properties of the Gaussian input state and the interferometer. This result is the equivalent to the result for Fock states (e.g. \cite{scheel2004permanents}), which provides the foundation for Boson sampling schemes with single photons.

We are interested in calculating the overlap of our Gaussian state $\hat{\rho}$ with the number state operator $\hat{\bar{n}}=\otimes_{j=1}^M \hat{n}_j$, where $\hat{n}_j=\ket{n_j}\bra{n_j}$ measures $n_j$ photons in output mode $j$. Typically this is,
\begin{equation}
\mathrm{Pr}(\overline{n})= \mbox{Tr}\left[\hat{\rho} \,\hat{\bar{n}}  \right]
\end{equation}
In our analysis we will use the phase space representation of quantum mechanics \cite{ferraro2005, Barnett_Radmore, schleich2011quantum}, similar to the approach used in \cite{Dodonov:1994p11102, RahimiKeshari:2015p11006}. Our formula of interest is now written as the overlap integral of the Q- and P-functions of the state and measurement operator respectively, 
\begin{equation}
\label{eq:phase-space}
\mathrm{Pr}(\overline{n})=\pi^M \int \mathrm{d} \boldsymbol{\alpha} Q_{\hat{\rho}}(\boldsymbol{\alpha}) P_{\bar{n}}(\boldsymbol{\alpha})\, 
\end{equation}
where $\mathrm{d}\boldsymbol{\alpha}=\prod_{j=1}^M \mathrm{d}\alpha_j \mathrm{d}\alpha_j^*$, $Q_{\hat{\rho}}(\boldsymbol{\alpha})$ is the Q-function representation of the Gaussian state  \cite{husimi_formal_1940} and $P_{\bar{n}}(\boldsymbol{\alpha})$ is the P-representation \cite{glauber_coherent_1963, sudarshan_equivalence_1963} of the number state operator.

An $M$-mode Gaussian state can be fully characterised by its $2M\times 2M$ covariance matrix $\sigma$ and a displacement vector $d$ \cite{Simon:1994p4225, ferraro2005}
\begin{equation}
\label{eq:sigma}
\sigma_{ij}=\frac{1}{2}\langle\{\hat{\zeta}_i, \hat{\zeta}^\dag_j\}\rangle-d_id^*_j \, ,\,\quad d_i=\langle\hat{\zeta}_i\rangle,
\end{equation}
where $\hat{\zeta}_i$ runs over all creation and annihilation operators $\hat{a}_j,\,\hat{a}_j^\dagger$ and we assume $d_i=0$ for this derivation (we discuss the case $d_i\neq 0$ in section \ref{sec:displacement}). Note, that $\sigma$ here corresponds to the measured modes of the system (i.e. at the output of an interferometer). If we do not measure a mode, then the corresponding rows and columns of that mode are removed from the covariance matrix and the state that remains is also a Gaussian state. From the covariance matrix $\sigma$, we can construct the Q-function of the state by convolving the corresponding Wigner function with another Gaussian function \cite{Barnett_Radmore}
\begin{equation}
Q_{\hat{\rho}}(\alpha)=\frac{1}{\sqrt{|\pi \sigma_Q|}}\exp\left[-\frac{1}{2}\alpha_\nu^\dag\sigma_Q^{-1}\alpha_\nu\right]\, ,
\label{eq:q-defn}
\end{equation}
where ${\sigma_Q=\sigma+\mathds{I}_{2M}/2}$ with $\mathds{I}_{2M}$ is the $2M\times 2M$ identity matrix and ${\alpha_\nu=[\alpha_1,\alpha_2...\alpha_M,\alpha_1^*,\alpha_2^*...\alpha_M^*]^t}$. 
The P-function of the $n$-photon number state $|n\rangle\langle n |$ is \cite{Gardiner_Zoller} 
\begin{equation}
P_n(\alpha)=\frac{e^{|\alpha|^2}}{n!}\left(\frac{\partial^2}{\partial \alpha \partial\alpha^*}\right)^n\delta(\alpha)\delta(\alpha^*)
\end{equation}
where $\delta(\alpha)$ is the two-dimensional Dirac-delta function ${\delta(\alpha)= \delta (\mathrm{Re}(\alpha))\delta (\mathrm{Im}(\alpha))}$. 
When we insert these into Eq. \eqref{eq:phase-space} and perform integration by parts we arrive at
\begin{equation}
\mathrm{Pr}(\overline{n})=\frac{1}{\overline{n}!\sqrt{|\sigma_Q|}} \left.\prod_{j=1}^M\left(\frac{\partial^2}{\partial\alpha^{}_j\partial\alpha_j^*}\right)^{n_j} \exp\left[\frac{1}{2}\alpha_\nu^t \textbf{A}\alpha_\nu\right] \right|_{\mathbf\alpha=0}\, ,
\label{eq:pr-exponent}
\end{equation}
where we have defined
\begin{equation}
\textbf{A}= \begin{pmatrix} 0 & \mathds{I}_M\\
			\mathds{I}_M & 0 \end{pmatrix} \left[\mathds{I}_{2M}-\sigma_Q^{-1}\right]\, . 
\label{eq:A_defn}
\end{equation}
We have switched from $\alpha_\nu^\dag$ in \eqref{eq:q-defn} to $\alpha_\nu^t$ in \eqref{eq:pr-exponent} ($\alpha_\nu^t=\alpha_\nu^\dag P$, with $P$ as a permutation matrix). We introduce $P$ only to reorder the vector $\alpha^\dag$ and thus simplify the final expression. 

In order to evaluate the expression in Eq. \eqref{eq:pr-exponent}, we expand the derivatives using Fa$\grave{\text{a}}$ di Bruno's formula, a higher order chain rule \cite{Comtet74}. For now, to stay in the typical Boson sampling framework, we restrict ourselves to measure either $n_j=\{0,1\}$ photons at each output mode (we will discuss higher photon numbers in a single output mode in section \ref{sec:manyphot}). For $N$ measured photons in total we have $2N$ derivatives ($\partial \alpha_j, \partial \alpha_j^* $ per photon) in Eq. \eqref{eq:pr-exponent}, each having an index $j$ (for $\alpha_j$) and $j+M$ (for $\alpha^*_j$). 
The expansion of the derivatives yields \cite{hardy2006combinatorics}
\begin{equation}
\frac{\partial^{2N}e^{\frac{1}{2}\alpha_\nu^t \textbf{A}\alpha_\nu}}{\prod_i^N\partial\alpha_i\partial\alpha_i^*}=e^{\frac{1}{2}\alpha_\nu^t\textbf{A}\alpha_\nu}\sum^{|\pi |}_{\substack{j=1 \\ \pi_j \in \{2N\} }} \left ( \prod^{|\pi_j|}_{\substack{k=1 \\ B_k \in \pi_j }} \frac{\partial^{|B_k|} \alpha_\nu^t \textbf{A}\alpha_\nu}{\prod^{|B_k|}_{\substack{l=1 \\ l \in B_k }}\partial\alpha_l^{(*)}}\, \right ),
\label{eq:partition}
\end{equation}
where the first sum runs over all partitions $\pi_j$ (where $|\pi|$ represents the number of partitions) of the set $\{\alpha_i^{(*)}=\alpha_i,\alpha_i^*\}$ (size $2N$), the first product over $B_k$ is over all $k$ blocks of the partition $\pi_j$ (the number of blocks of $\pi_j$ is $|\pi_j|$). The partial derivative is formed from the size of the block $|B_k|$ (the number of indices contained within $B_k$) which gives the order of the derivative and is differentiated with respect to the elements of that block, the $\alpha_l$ or $\alpha^*_l$.

Thus, the expansion of the derivatives can be related to the different partitions of the set of photon indices. To illustrate this point, we consider the case when a single photon is detected in both mode 1 and 2, and thus we have to find all partitions of the set of indices $\{\alpha_1, \alpha_1^*,\alpha_2,\alpha_2^*\}$. 
One such partition, $\{\alpha_1\},\{\alpha_1^*,\alpha_2,\alpha_2^*\}$, corresponds to the term in the derivative expansion 
\begin{equation}
\frac{\partial \alpha_\nu^t\textbf{A}\alpha_\nu}{\partial \alpha_1}\frac{\partial^3\alpha_\nu^t\textbf{A}\alpha_\nu}{\partial \alpha_1^* \partial\alpha_2\partial\alpha_2^*}\, .
\label{eq:partition2}
\end{equation}

When calculating the derivatives of $\alpha_\nu^t\textbf{A}\alpha_\nu$ in Eq. \eqref{eq:pr-exponent}, we find that, as it is a quadratic function of $\alpha_\nu$, all derivatives of third order or higher vanish. In addition, since we evaluate the derivatives at $\alpha_\nu=0$, all derivatives of first order also vanish. We are therefore only left with the partitions where the $2N$ elements are sorted into $N$ sets, each of size 2. This means that, in the above formalism, for $2N$ variables, ${|\pi_j|=N \, \forall j} $, ${|B_k|=2\,\forall \, k}$ and the number of partitions is $(2N-1)!!$, where $(.)!!$ denotes the double factorial\footnote{In the case, where the argument of the double factorial is even, $(2N)!!$ the product runs over all even numbers less than or equal to $2N$.}, the product over all odd numbers less than or equal to $2N-1$. 

These partitions (of $2N$ numbers into $N$ blocks of size 2) can be interpreted as permutations of the $2N$ photon indices, which can be written in a vector $\mu_j$. For each partition, the blocks are ordered with respect to their smallest element (lowest to highest) and the numbers within a block are also ordered in increasing size. In terms of the permutation vector $\mu$, these conditions can be written as 
\begin{align}
1)&\, \mu_j({2k-1})<\mu_j({2k})  \nonumber\\
2)&\, \mu_j({2k-1})<\mu_j({2k+1)\,}. \nonumber
\end{align}
for $k=1,...,N$. The set of permutations that satisfy these conditions are known as the perfect matching permutations (PMP) \cite{callan2009combinatorial} and there are $(2N-1)!!$ such permutations (or partitions)
\footnote{E.g. For $N=2$ photons detected in modes 3 and 4 of a $M=4$ modes unitary we have to consider the set of indices: ${\{3,4,7,8\}}$. The number of PMP is ${(2N-1)!!=3}$. The partitions (and permutations) are then
\begin{eqnarray*}
&\pi_1=\{34\}\{78\},\ 
\pi_2=\{37\}\{48\}, \ 
\pi_3=\{47\}\{38\}\\
&\mu_1 = 3,4,7,8 \,\
\mu_2 = 3,7,4,8 \, \
\mu_3 = 4,7,3,8
\end{eqnarray*}
}.

With this definition, we are now able to write down the final result for Eq. \eqref{eq:pr-exponent}
\begin{equation}
\mathrm{Pr}(\overline{n})=\frac{1}{\overline{n}!\sqrt{|\sigma_Q|}}\sum^{(2N-1)!!}_{ \mu_j \in \{PMP\}} \prod_{k=1}^N \textbf{A}_{{\mu_j(2k-1),\mu_j(2k)}}\, ,
\label{eq:probability-evaluated}
\end{equation}
The indices of the measured photons' position, stored in $\mu$, define $\mathbf{A}_S$, a submatrix of $\mathbf{A}$. The sum in \eqref{eq:probability-evaluated}, over all PMP of $\mathbf{A}_S$, is exactly the hafnian of that matrix, as defined by Caianiello \cite{Caianiello:1953p12510, Caianiello73}. As such, we are able to write down a closed-form expression that connects the probability to measure a specific output pattern $\bar{n}$ from any Gaussian state with the hafnian matrix function \cite{hamilton_gaussian_2016}
\begin{equation}
\mathrm{Pr}(\overline{n})=\frac{1}{\overline{n}!\sqrt{|\sigma_Q|}}\mathrm{Haf}(\textbf{A}_S)\, .
\label{eq:hafnian-probability}
\end{equation}
This formula comprises the basis for a truly Gaussian Boson Sampling protocol due to the nature of the hafnian function. We will discuss the hafnian in the next section. 

\begin{figure}
\includegraphics[width=.8\columnwidth]{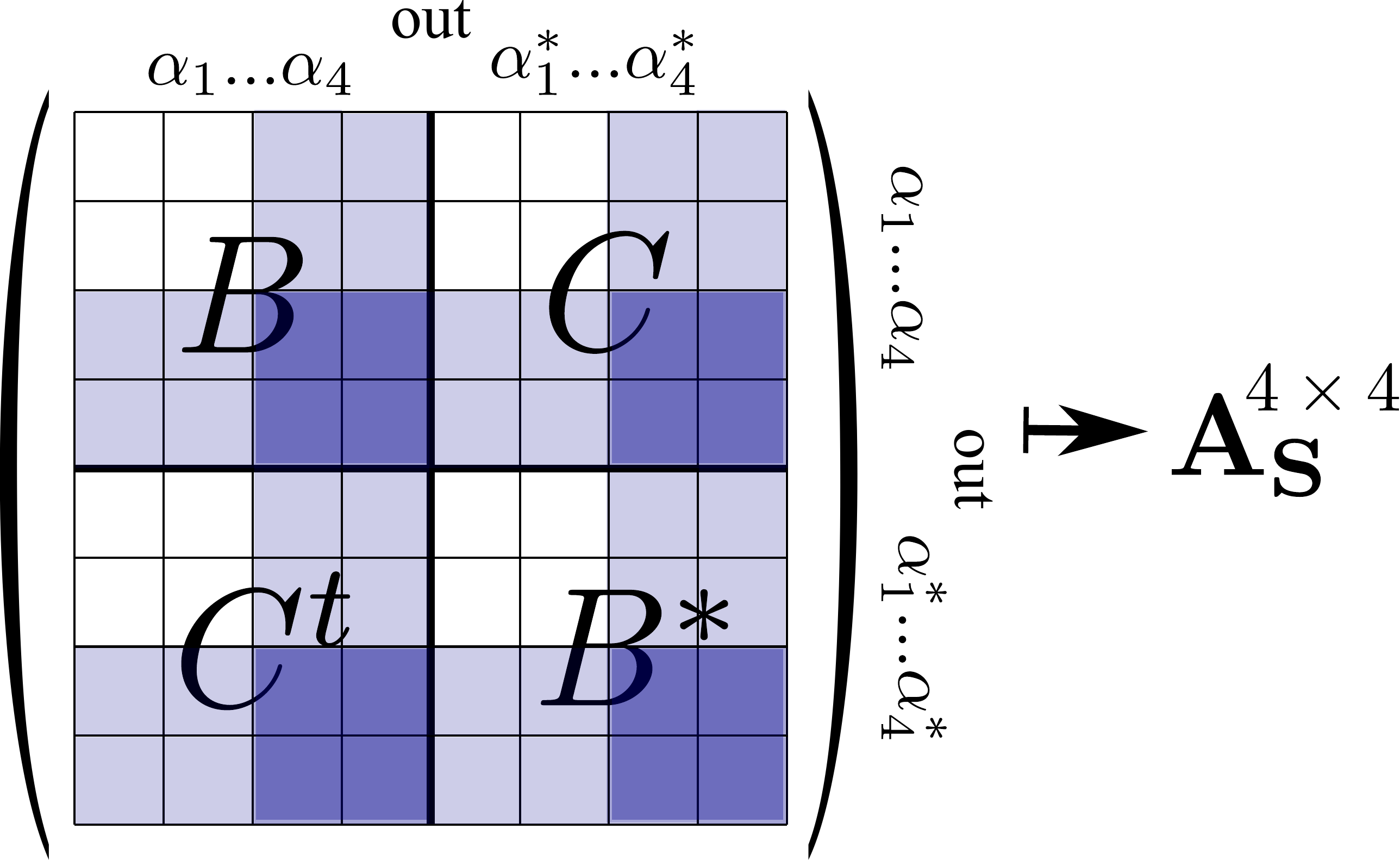}
\caption{Construction of the submatrix $\mathbf{A}_S$ from the state matrix $\mathbf{A}$ for two photons measured in last two output modes 3 and 4 of an M=4 mode interferometer. Contrary to the Fock Boson Sampling schemes, the selection of the matrix entries in $\mathbf{A}_S$ is independent of the input state and only depends upon the output photon pattern, $\bar{n}$. For details see text.
}
\label{fig:gaussian-submatrix}
\end{figure}

As $\mathbf{A}$ is a symmetric matrix of dimension $2M \times 2M$, due to the structure of the initial covariance matrix, it can be divided into four blocks of dimension $M \times M$, as indicated in figure \ref{fig:gaussian-submatrix}. The structure of $\mathbf{A}$ is a combination of the squeezed and thermal contributions present in the state. However, if we only have squeezed light present in our state, then $\mathbf C=0$ and $\mathbf B\neq 0$ and if we only have thermal light then the opposite is true, $\mathbf B=0,~\mathbf C\neq 0$. For the latter case, our formalism reproduces the results for thermal states derived in \cite{RahimiKeshari:2015p11006} by using a matrix identity for the hafnian \cite{Minc78} [cf. Eq. \eqref{eq:permanent-hafnian-relation}].
The construction of the submatrix $\mathbf{A}_S$ depends, in contrast to standard Boson sampling schemes, only on the measured output pattern (compare figures \ref{fig:AABS}(a) and \ref{fig:gaussian-submatrix}). Any detected $N$-photon event then selects a $2N\times 2N$ submatrix, where a detected photon in mode $j$ selects the columns $j$ and $j+M$ of $\mathbf{A}$, and the rows with the same indices. This is illustrated for a two photon example
\footnote{In this case photons detected in modes 3 and 4, from $M=4$ overall modes, select the $4\times 4$ submatrix
\begin{equation}
\nonumber
\mathbf{A}^{4\times 4}_S=
\begin{pmatrix}
A_{33} & A_{34} & A_{37} & A_{38}\\
A_{43} & A_{44} & A_{47} & A_{48}\\
A_{73} & A_{74} & A_{77} & A_{78}\\
A_{83} & A_{84} & A_{87} & A_{88}							 
\end{pmatrix}\, .
\end{equation}
} by the blue bars in figure \ref{fig:gaussian-submatrix}. 

\subsection{Multiple photons in the same mode}
\label{sec:manyphot}

In the above derivation we restricted ourselves to the case where we only detect $n_j=\{0,1\}$ photons per output mode, however the formalism of Eq. \eqref{eq:hafnian-probability} is not limited to this case. To consider the case of having more than one photon per output mode we have to adapt the submatrix that we sample from. Consider the simplest example, a single-mode system. The system matrix $\mathbf{A}$ is given by
\begin{equation}
\textbf{A}=\begin{pmatrix} A_{11} & A_{12}\\
			   A_{21} & A_{22} \end{pmatrix}\, .
\end{equation}
If we now consider a two-photon detection event in this mode then Eq. \eqref{eq:pr-exponent} is given by
\begin{equation}
\mathrm{Pr}(n_1=2)=\frac{1}{\sqrt{|\sigma_Q|}}\left.\frac{1}{2!}\frac{\partial^2}{\partial\alpha_1^2}\frac{\partial^2}{\partial\alpha_1^{*^2}}e^{\frac{1}{2}\alpha_\nu^t\mathbf{A}\alpha_\nu}\right|_{\mathbf \alpha=0}\, .
\end{equation}
Terms like this are not covered directly by the calculation of the hafnian. We can circumvent this problem by artificially ``moving'' this photon to another ``psuedo-mode'', and form a new matrix $\mathbf{A}'$ by repeating the corresponding rows and columns of $\mathbf{A}$, i.e. we write
\begin{equation}
\begin{aligned}
&\mathrm{Pr}(n_1=1,n_2=1)=\\
&\quad\quad\frac{1}{2!\sqrt{|\sigma_Q|}} \left.\frac{\partial}{\partial\alpha_1}\frac{\partial}{\partial\alpha_1^*}\frac{\partial}{\partial\alpha_2}\frac{\partial}{\partial\alpha_2^*}e^{\frac{1}{2}\alpha_\nu^t\mathbf{A'}\alpha_\nu}\right|_{\mathbf \alpha=0}\, ,
\end{aligned}
\end{equation}
where we have defined $\mathbf{A}'$ as a new matrix constructed as
\begin{equation}
\textbf{A}'
		=\begin{pmatrix} 	A_{11} & A_{12} & A_{11} & A_{12} \\
							A_{21} & A_{22} & A_{21} & A_{22} \\
							A_{11} & A_{12} & A_{11} & A_{12} \\
							A_{21} & A_{22} & A_{21} & A_{22} \end{pmatrix}\, .
\end{equation}
This can be repeated for each extra photon in that mode, such that there is always one mode per photon and $\mathbf{A}'$ is $2N \times 2N$ matrix. Note, that $\mathbf{A}'$ is not a proper quantum covariance matrix. We only define it as a way to use the hafnian expression for higher order photon detection events.


\subsection{Non-zero displacement}\label{sec:displacement}

Finally, we analyse the situation where we consider a non-zero displacement in our state, i.e. we allow for $\langle\hat{\zeta}_j\rangle=d_j\neq 0$ in Eq. (\ref{eq:sigma}). In this case, the Q-function for a displaced, multimode Gaussian state (squeezed and thermal contributions) is given by
\begin{equation}
\begin{aligned}
&Q(\alpha,\alpha^*)= \\
&\frac{1}{\sqrt{|\sigma_Q|}} \exp \left[ -\frac{1}{2}(\alpha_{\nu} - d_{\nu})^\dag \sigma_Q^{-1} (\alpha_{\nu}-d_{\nu})\right]\, .
\end{aligned}
\end{equation}
Expanding the exponent yields
\begin{equation}
\begin{aligned}
-\frac{1}{2}(\alpha_\nu - d_\nu)^\dag \sigma_Q^{-1} &(\alpha_\nu-d_\nu) \\
&= -\frac{1}{2}d_\nu^\dag \sigma_Q^{-1} d_\nu -\frac{1}{2}\alpha_\nu^\dag \sigma_Q^{-1} \alpha_\nu +F\alpha_\nu\, ,
\end{aligned}
\end{equation}
where we defined $F= d_\nu^\dag \sigma^{-1}_Q$. Inserting this into Eq. \eqref{eq:phase-space} (or \eqref{eq:pr-exponent}), we arrive at
\begin{equation}
\begin{aligned}
\Pr(\bar{n})&= 
\frac{\exp\left[-\frac{1}{2}d_{\nu}^\dag \sigma_Q^{-1} d_{\nu}\right]}{\bar{n}!\sqrt{|\sigma_Q|}}\\
&\times\left. \prod^{M}_{j=1} \left(\frac{\partial^2}{\partial \alpha_j \partial \alpha_j^*}\right)^{n_j}\hspace{-3mm}\exp\left[\frac{1}{2}\alpha_\nu^t\mathbf{A}\alpha_\nu+ F \alpha_\nu\right]\right|_{\mathbf\alpha=0} \, .
\end{aligned}
\end{equation}
As $F\alpha_\nu$ is a linear function of $\alpha_\nu$ we have extra, non-zero terms in the expansion of the derivatives \eqref{eq:partition}, when compared to the squeezing only case \eqref{eq:partition2}. That means that we now have first order terms appearing in the expansion of the derivatives in Eq.~\eqref{eq:partition}. For example, it is now possible that partitions of the form $\{\alpha_1\},\{\alpha_1^*\}, \{\alpha_2\},\{\alpha_2^*\}$ or $\{\alpha_1\},\{\alpha_1^*\},\{\alpha_2,\alpha_2^*\}$ will contribute to the overall probability. These partitions, respectively, lead to terms in the expansion of the derivatives
\begin{equation}
\nonumber
\begin{aligned}
&\frac{\partial F\alpha_\nu}{\partial \alpha_1} \frac{\partial F\alpha_\nu}{\partial \alpha^*_{1}} \frac{\partial F\alpha_\nu}{\partial \alpha_{2}} \frac{\partial F\alpha_\nu}{\partial \alpha^*_{2}}= F_1  F_{1+M} F_2 F_{2+M}\quad\quad\text{and} \\
&\frac{\partial F\alpha_\nu}{\partial \alpha_1} \frac{\partial F\alpha_\nu}{\partial \alpha^*_{1}} \frac{\partial^2 \alpha^\dag_\nu A\alpha_\nu}{\partial \alpha_2 \partial \alpha^*_{2}} = F_1 F_{1+M} A_{2,2+M}
\end{aligned}
\end{equation}
(and we have ignored contributions that evaluate to zero at $\alpha_\nu=0$). 
Re-examining Eq.~\eqref{eq:partition}, we now have a total number of partitions
\begin{equation}
|\pi|=\sum^N_{k=0} {{2N}\choose{2k}} (2(N-k)-1)!!
\end{equation}
instead of $(2N-1)!!$.
The individual partitions are formed by first taking $2k$ of the $2N$ variables, to give $2k$ single-index partitions and $N-k$ double-index partitions. This subset gives us a product of the first order terms $F_j$, corresponding to those indices within the subset. The remaining $2N-2k$ indices give us a submatrix of $A$, and we calculate the hafnian of this submatrix. We can write each partition of the $2N$ numbers as
\begin{equation}
\pi_j =  \bigcup_{l=1}^{2k} B^{1}_{l} \bigcup_{l'=1}^{2N-2k} B^{2}_{l'}
\end{equation}
where $B^1$ are the single-index blocks of $\pi_j$ and $B^2$ are the blocks of size 2 (as we had before). This leads to a modified expression for the probability of a photon pattern, akin to Eq. \eqref{eq:hafnian-probability}, 

\begin{equation}
\begin{aligned}
&\mathrm{Pr}(\bar{n}) =\frac{e^{-\frac{1}{2}d_{\nu}^\dag \sigma_Q^{-1} d_{\nu}}}{\bar{n}!\sqrt{|\sigma_Q|}} \\
&  \times \sum^{|\pi |}_{\substack{j=1 \\ \pi_j \in \{2N\} }} \left[  \left ( \prod^{|B^1_j|}_{\substack{k =1 \\ B_j^1 \in  \pi_j}} F_k \right ) \mathrm{Haf}(A_{B_j^2}) \right]
\\
 &=\frac{e^{-\frac{1}{2}d_{\nu}^\dag \sigma_Q^{-1} d_{\nu}}}{\bar{n}!\sqrt{|\sigma_Q|}} \Bigg [ \mathrm{Haf}(A_S)  \\
&  + \sum_{j_1,j_2 ,j_1\ne j_2} F_{j_1} F_{j_2} \mathrm{Haf}(A_{S - \{j_1,j_2\}}) + ... + \prod^{2N}_j F_j \Bigg ]  \\
\end{aligned}
\label{eq:squ_coh_light}
\end{equation}
where the first sum is over all partitions of the set  of $2N$ indices, the product is over all indices in the blocks $B^1_j$ and the remaining indices in blocks $B^2_j$ form $A_{B^2_j}$, a submatrix of A, which we then take the hafnian of. 

We can give an interpretation to the terms in Eq. \eqref{eq:squ_coh_light}. The first term in the sum can be identified as the contribution where all the photons come from the covariance matrix (squeezed and thermal light) and none from displacement operator. The last term only contains the contributions from the displacement operators, i.e. when all the photons come from the coherent state. The intermediate terms mix photons from both the squeezed, thermal and coherent contributions of the state.

In the case where we only have coherent light ($\sigma_Q = \mathds{I}$), Eq. \eqref{eq:squ_coh_light} reduces to
\begin{equation}
\mathrm{Pr}(\bar{n}) = \frac{e^{-\frac{1}{2}d_{\nu}^\dag \sigma_Q^{-1} d_{\nu}}}{\bar{n}!\sqrt{|\sigma_Q|}} \prod^{2N}_{j=1} F_j = \frac{e^{-\sum_j |d_j|^2} }{\bar{n}!}\prod_{j=1}^{N} |d_j|^{2n_j}\, ,
\end{equation}
as expected \cite{Barnett_Radmore}. Depending on the squeezing and displacement levels in our state, the weights of the respective contributions vary i.e. for an almost purely squeezed state, the first term will dominate the other terms and for a large displacement, the last term will dominate the photon counting probability.


\section{Construction of GBS with squeezed states}\label{sec:GBS}
\begin{figure}
\includegraphics[width=.8\columnwidth]{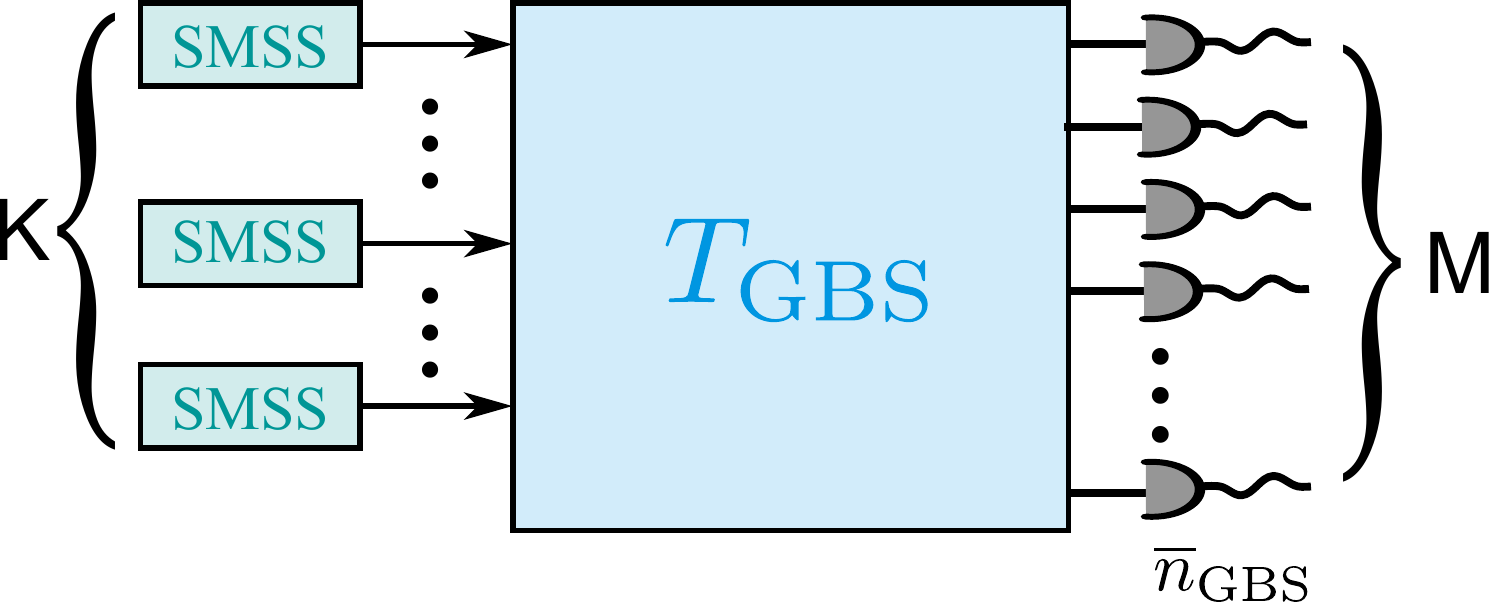}
\caption{Schematic of the GBS protocol. We send $K$ single mode squeezed states into a Haar random interferometer $T_\mathrm{GBS}$ of size $M$ and sample the output photon distribution $\bar{n}_\mathrm{GBS}$ at the end.}
\label{fig:GBS}
\end{figure}
In this section we develop the protocol Boson sampling from a Gaussian state. We start by describing the main requirements for a Gaussian Boson sampling protocol, and in subsequent sections we comment on the details of such a protocol, including approximate GBS. 

The main requirement for Fock Boson sampling protocols is the computational complexity of the underlying matrix function, the permanent, which is in the \textsf{\#P} complexity class. The hafnian, also in the \textsf{\#P}-class \cite{Valiant:1979p11225}, is a more general function than the permanent, as the hafnian counts the number of perfect matchings in a general, undirected graph whereas the permanent is restricted to a bipartite graph.  
This is encapsulated in the formula
\begin{equation}
\mathrm{Perm}(G)=\mathrm{Haf}\left[\begin{pmatrix} 0 & G \\
												   G^t & 0 \end{pmatrix}\right]\, .
\label{eq:permanent-hafnian-relation}
\end{equation}
where we can express the permanent of a matrix $G$ in terms of the hafnian \cite{Minc78}. 

Having discussed this necessary requirement for a Boson sampling problem, we proceed to construct the GBS protocol based on squeezed states. We use squeezed states as it is known that thermal states can be approximated in BPP$^\mathrm{NP}$  \cite{chakhmakhchyan_perm_therm, RahimiKeshari:2015p11006}, a complexity class  easier than \#P.

We depict the physical setup of the protocol in Fig. \ref{fig:GBS}, where $K$ single mode squeezed states enter a linear interferometer $\mathbf T_\mathrm{GBS}$ and at the output we measure all $M$ modes of the system and record all photo-counts. This choice of squeezing and linear transformation leads to $\mathbf B\neq 0$ and $\mathbf C=0$ in the overall system matrix $\mathbf{A}$ (see Fig. \ref{fig:gaussian-submatrix}). For this scheme, the matrix $\mathbf{A}$ is defined by the input state and the interferometer $\mathbf T_\mathrm{GBS}$. The single mode squeezed states in our system are described by the matrix
\begin{equation}
S=\begin{pmatrix} \bigoplus_{j=1}^M\mathrm{cosh} \,r_j & \bigoplus_{j=1}^M\mathrm{sinh}\, r_j\\
						\bigoplus_{j=1}^M\mathrm{sinh} \,r_j & \bigoplus_{j=1}^M\mathrm{cosh}\, r_j\end{pmatrix}\, ,
\end{equation}
where $r_j$ is the squeezing parameter of the single mode squeezed states in the $j$-th mode and ${\bigoplus_{j=1}^M x_j=\mathrm{diag}(x_1,x_2,...x_M)}$, a direct sum of numbers, yielding a diagonal matrix. Note that $r_j=0$, for $M-K$ entries, corresponds to a vacuum state input. Then, the covariance matrix at the output of the interferometer is given by
\begin{equation}
\sigma=\frac{1}{2}\begin{pmatrix}T_\mathrm{GBS} & 0\\
								0 & T^*_\mathrm{GBS}\end{pmatrix}SS^\dagger
															\begin{pmatrix}T^\dagger_\mathrm{GBS} & 0\\
																			0 & T^t_\mathrm{GBS}\end{pmatrix}
\end{equation}
and $\mathbf{A}$ in Eq.~\eqref{eq:A_defn} is calculated to be $\mathbf{A}=\mathbf{B\oplus B^*}$, with
\begin{equation}
B=T_\mathrm{GBS}\left(\bigoplus_{j=1}^M\mathrm{tanh}\,r_j\right)T^t_\mathrm{GBS}\, . \label{eq:B_defn}
\end{equation}
It is easy to show that the hafnian of a direct sum, as in $\mathbf{A}$, can be written as the product of the hafnians of the two submatrices. Thus, we can simplify Eq. \eqref{eq:hafnian-probability} to 
\begin{equation}
\mathrm{P}(\overline{n})=\frac{1}{\sqrt{|\sigma_Q|}}|\mathrm{Haf}(B_S)|^2\, 
\label{eq:hafnian_squared}
\end{equation}
where to construct this matrix we have restricted ourselves to the measurement outcome of $n_j=\{0,1\}$ per mode. As $\mathbf B_S$ is a submatrix of $\mathbf A_S$, its construction is obtained by keeping the intersection of the rows and columns where a photons was measured, a single index per photon. $\mathbf B_S$ will be an even-sized matrix, as, physically, this corresponds to measuring an even number of photons from the multimode squeezed state. The probability to measure an odd-number of photons from such a state is always zero. Note, that in the case of odd $N$, Eq.~\eqref{eq:hafnian-probability} still applies, but the identity \eqref{eq:hafnian_squared} is invalid. 

Due to the intrinsic complexity of the hafnian the complexity of GBS in the exact case is ensured. However, this does not guarantee the complexity for an approximate Gaussian Boson sampling protocol, which we discuss next. 


\subsection{Complexity of displacement contributions and multiple photons in the same mode }

In this section we comment on the complexity of the two other instances of the GBS expression, that of multiple photons in the same mode and of the contribution of displaced light. 

As shown in the previous section, we can incorporate the measurement result of multiple photons in the same mode by modifying the matrix $\mathbf A$. The extra photons can be included by repeating the rows and columns of the original matrix to generate an extended matrix. These extra rows/columns do not increase the rank of the matrix $\mathbf A$ and thus do not increase the complexity of calculating the output pattern in the way that detecting a photon in another mode would. While this method allows us to write the expression using the hafnian, a more computationally efficient method to incorporate multi-photon events was described by Kan \cite{Kan2008}.

The complexity of measuring photons from a displaced states is in the P complexity class, as the output state can be written as a vector of displacement amplitudes and the probability of photon numbers in each mode is independent of each other. This is in contrast to squeezed or thermal states, where the complexity arises from the correlations between modes. From Eq.~\eqref{eq:squ_coh_light}, the complexity of the combination of squeezed and displaced light still comes from the squeezed light (the hafnian terms) and therefore displaced light does not increase the complexity of the problem. 

%
\section{Approximate GBS}\label{sec:approx}

In this section we present two main heuristic arguments for approximate GBS to be in \#P. The first idea is similar to the proof of AABS \cite{aaronson_computational_2011} and uses a key result of theirs, which is to hide the matrix we wish to sample within a larger unitary transformation. The second is unique to GBS, which is to use the degrees of freedom of the input state to control the matrix we sample from, allowing us to reduce the size of the overall unitary matrix.

Before we present the ideas of approximate GBS, we briefly recall the main arguments of AA that approximate AABS is a \#P-hard problem. 
The approximate AABS problem $|\text{GPE}|^2_\pm$ states that given a matrix $X\in \mathbb{C}^{n\times n}$ of independent and identically distributed (i.i.d.) complex normal entries and error bounds $\epsilon, \delta$, the estimation of the permanent $|\mathrm{Perm}(X)|^2$ up to an additive error $\pm\epsilon\, n!$ with a success probability of $1-\delta$ for any possible $X$ takes a time polynomial in $(n,1/\epsilon,1/\delta)$. 
The main requirement that the Boson Sampling computer has to fulfil in this instance is that it is ``robust", meaning that if a small fraction $\epsilon$ of all events are ``badly wrong", the remaining $1-\epsilon$ results are still valid to encode the Boson Sampling scheme.

If we suppose that an approximate Boson Sampling computer works this way, we can use the robust encoding to prevent a classical adversary from corrupting our sampling. The procedure to show that approximate AABS up to an additive error is hard uses the fact that we hide the interesting probability (i.e. the sampling of a specific $\mathbf T_S$) among all the other random outputs of our Boson Sampling scheme. The solution that AA propose is to choose the $M \times M$-dimensional interferometer matrix $\mathbf T$ according to the Haar measure. Then, any sufficiently small submatrix is, in variation distance, close to a matrix whose entries are independent and identically drawn from the complex normal distribution. 
This means that the adversary will not know in which instance we are interested in and therefore cannot corrupt the result, on average. 

The sampling from such a device is random in the sense that we cannot predict the output pattern of $N$ photons, even if the same input state is used. This choice fulfils the robustness criterion and the need to hide the interesting sampling probability in a multitude of other possible output patterns. Then, using Stockmeyer's algorithm \cite{stockmeyer_complexity_1983}, AA show that $|\text{GPE}|^2_\pm$ is in $\textsf{BPP}^{\textsf{NP}^\mathcal{O}}$, where $\mathcal{O}$ is an oracle for approximate AABS. 
If there is a classical algorithm to simulate $\mathcal{O}$ then the polynomial hierarchy will collapse, having severe consequences for the computational complexity theory. 

While $\textsf{BPP}^\textsf{NP}$ is enough to claim that Boson Sampling is not classically ``simple", it remains an open question, whether approximate AABS is indeed in \textsf{\#P}. Nevertheless, AA provide evidence in the form of two conjectures, the \textit{permanents-of-gaussians} conjecture, which says that estimating the permanent up to multiplicative error $\text{GPE}_\times$ is in \textsf{\#P}. The second \textit{permanent-anti-concentration} conjecture implies a polynomial-time equivalence of the sampling up to additive error $|\text{GPE}|^2_\pm$ and the sampling up to multiplicative error $\text{GPE}_\times$. If these conjectures hold, this would mean that ${\textsf{P}^\textsf{\#P}=\textsf{BPP}^\textsf{NP}}$, unless approximate Boson Sampling is in \textsf{\#P}.

\subsection{First argument for approximate GBS}\label{app_gbs_1}

This argument follows the same (heuristic) steps of AA. We also assume a ``robust" encoding and strive to hide the ``interesting" probability among all outputs of our Gaussian Boson Sampling computer.

In GBS, we sample a matrix of the form $T\Gamma T^\dag$, Eq.~\eqref{eq:B_defn} where $\Gamma$ describes the input squeezed states and $T$ is a unitary matrix that describes the interferometer that the photons travel through. 
From the results of AA, we can ``hide'' the matrix $X$ in $T$ and, if we choose the shape of our input squeezed state to be $\Gamma = \mathds{I}_N \oplus 0_{M-N}$, then we can associate the ``interesting outcome'' with the matrix $XX^t$. Then, analogously to ${|\mathrm{GPE}|_\pm^2}$ in AABS, one can formulate a Gaussian Hafnian estimation problem $|{\mathrm{GHE}|_\pm^2}$, where the task is to estimate $|\mathrm{Haf}(XX^t)|^2$ of a symmetric matrix with random entries up to an additive error. 

One difference between GBS and other Boson sampling protocols is that in the former the number of photons is not fixed (see sections \ref{sec:GBS_events} \& \ref{sec:number_dist}  below) because of the nature of Gaussian states. We can restrict our device to a fixed photon number at a cost polynomial in that number. This means we can focus on the same class of output states in AABS. Based on these arguments and given an approximate GBS oracle $\mathcal O$, a combination of Stockmeyer's algorithm and Markov's inequality should yield that $|\mathrm{GHE}|_\pm^2$ is in $\textsf{BPP}^{\textsf{NP}^\mathcal{O}}$.

As in AA, we leave open the final proof that approximate GBS is $\textsf{\#P}$ hard. Yet, as we already stated in \cite{hamilton_gaussian_2016}, we conjecture that approximating the hafnian to a multiplicative error $\mathrm{GHE}_\times$ is in \textsf{\#P}, i.e. a \textit{hafnian-of-gaussians} conjecture. This is a generalisation of AAs \textit{permanent-of-gaussians} conjecture, which we believe is justified, as the computation of a permanent can be reduced to the computation of a hafnian. Furthermore, we conjecture, analogously to AA, that the two problems $|\mathrm{GHE}|_\pm^2$ and $\mathrm{GHE}_\times$ are polynomial-time equivalent, which we infer from the very similar structure of both the permanent and the hafnian. If these two conjectures hold, then either approximate GBS is in the \textsf{\#P} complexity class or the polynomial hierarchy collapses.

\subsection{Second argument for approximate GBS}\label{app_gbs_2}

We now introduce a new perspective to devise an approximate GBS protocol, unique to our system. This is motivated by the fact that we have additional control over our system, namely that we can alter the initial input state by the squeezing parameters of the individual single mode squeezed states, a property which is not present in either AABS or SBS. 

In GBS the matrix that we sample from is given by Eq.~\eqref{eq:B_defn}.
This construction, if we can control both $\mathbf T_{GBS}$ and each $r_j$, means that we can generate {\it any} (rescaled) symmetric matrix $\mathbf B$ by use of the Autonne-Takagi decomposition \cite{Horn_Johnson}. This is a type of singular-value decomposition and factorises a complex, symmetric matrix into a unitary matrix and a diagonal matrix of positive numbers  (in the range $[0,\infty]$.) This means we can adjust our hafnian problem (from the previous section) to estimate $|\haf(X)|^2_\pm$, the hafnian of a symmetric matrix of random numbers from complex normal distribution (rather than $XX^t$).  
This scheme changes the requirements for hiding our matrix of interest. If we require $X$ to be a matrix of complex normal numbers, we can hide this in a larger matrix $X'$, also of complex normal numbers, not a unitary matrix as before. 
We can calculate this larger matrix using the Autonne-Takagi decomposition, which can be done exactly with no approximations needed. This is our `hiding lemma', with the remaining question is how large does the matrix $\mathbf T$ need to be to hide a $N\times N$ submatrix within it. 
Here we will conjecture that it needs to be only a {\it linear} factor of $N$, $M=O(N)=\kappa N$, and not a quadratic relationship as in AABS (and our first argument for approximate GBS in the previous section). This is because the number of potential output photon number patterns still increases exponentially with the number of photons and the error that the adversary adds to the device will be spread across these outcomes, as in AABS. As before, in the previous section, we can restrict ourselves to the output states within a set of total photon number ($2N$) at only polynomial cost (as shown below).

Therefore if we want to sample from a particular matrix $X'$, we find the decomposition $X'=UDU^t$ and then rescale it by $\sqrt{2}\lambda_\mathrm{max}$, where $\lambda_\mathrm{max}$ is the maximum singular value of $X$. This is because the $\tanh r_j$ that appear in the diagonal matrix of Eq.~\eqref{eq:B_defn}, can only take values between $[0,1]$ ($r\in[0,\infty]$). The rescaled matrix $D/(\sqrt{2}\lambda_\mathrm{max})$ corresponds to the set $\{\tanh r_j\}$, the squeezing parameters of the initial input states and $U=T_{GBS}$, the interferometer that this state enters. 

The next step is to proceed through the same analysis as AA to show that we can bound the error between the actual distribution and the output from our GBS device. As the size of our state space is exponentially large, an adversary could not corrupt enough events to make the device fail, according to the constraint above. We could then use the same arguments as above to show
that a classical algorithm for GBS would imply that the polynomial hierarchy collapses to the third level. 

Operating a GBS device in this regime reduces the size of the interferometer needed, a substantial improvement in the implementation of Boson sampling experiments. The experimental challenge of this is the necessary control over the squeezing parameters of the individual squeezed input states.


\section{Further requirements}\label{sec:requirements}

In the previous section, we outlined our arguments that the approximate GBS problem is also in the \textsf{\#P} complexity class. However, there are several aspects unique to GBS that must be satisfied to guarantee that the sampling is also complex. We now comment on those, as well as on optimal experimental parameters.

\subsection{Number of single mode squeezed states}

For permanents and hafnians it is known that the matrix rank determines the complexity of the computation \cite{Barvinok96, Kan2008}. The rank of the matrix that we sample in GBS, Eq. (\ref{eq:B_defn}), is determined by the number of independent single mode squeezed states. This means that if we want to sample $N$ photons, then we have to pump at least $K=N$ input modes with single mode squeezed states to saturate the complexity. Therefore, we require $K\geq N$ single mode squeezed states at the input of the interferometer. Note that this assumes we are working towards the approximate GBS in the first regime. The second regime will require all modes to be pumped with different squeezing parameters to exactly sample the correct matrix. 


\subsection{Dilute sampling}

In Sec. \ref{sec:hafnian} and Sec. \ref{sec:GBS} we required that we measure only $n_j=\{0,1\}$ in each output mode to avoid the repetition of rows and columns in the $\mathbf B_S$ matrix. The reason is that these `repeated' photons do not increase the rank of the sampled matrix, and thus the complexity of the Boson sampling problem \cite{Kan2008}. Therefore, we have to show that the probability to measure more than one photon in an output mode can be made sufficiently small.

Consider $N$ single mode squeezed states at the input, each with a mean number of $1$ photon ($\sinh^2 r=1$). Then, if we consider an interferometer of size $M=N^2$ that is balanced (all entries are of similar size), we have at the output, a mean number of $\tfrac{1}{N}$ photons per mode. This is due to the interferometer distributing all photons equally on average among the output modes, which a Haar random unitary can provide due to the intrinsic randomness of the Haar measure. If we now examine a single output mode of such a system and trace over all other modes, we obtain, approximately, a thermal state with a mean photon number $\langle n \rangle \approx \tfrac{1}{N}$. As a rule-of-thumb guide to the concentration of photons within the setup, we calculate the ratio between the probability of two-or-more photo-counts versus the probability of one photo-count for a single-mode thermal state,

\begin{equation}
\frac{\sum^\infty_{n_j\ge2}\mathrm{Pr}(n_j)}{\mathrm{Pr}(n_j=1)} = \frac{1}{N}\, .
\end{equation}

Due to this finite, but low, probability to measure two or more photon events in the same output mode, we require photon number resolution for our detectors. Yet, as the higher order coincidences have a very low probability of occurring, a low photon number resolving capability is enough to faithfully exclude higher order events in a single channel. This is the same requirement that SBS has in the heralding part of the scheme.

A similar analysis for the case in the second argument of approximate GBS, where $\bar{n}=1/\kappa$, yields that the above ratio is also ${ 1/\kappa}$. This means that the number of modes to photon number must be sufficiently large to ensure the former ratio is low enough to satisfy dilute boson sampling requirement. 


\subsection{Valid GBS events}\label{sec:GBS_events}

In Fock Boson sampling experiments, such as AABS, a fixed number of photons enter and exit the linear interferometer $\mathbf T$. That means that these experiments sample from the family of photon patterns with $N$ photons $\{P_N\}$
\begin{equation}
\{p_1,p_2,...p_{C_N}\}_N=\{P_N\}\, ,
\end{equation}
where $p_j$ is the probability of a particular pattern and $C_N={{M}\choose{N}}$ is the number of possible patterns of $N$ single photons in $M$ modes. We discard configurations with more than one photon in any output mode and thus $\sum_j p_j < 1$.

As we use Gaussian states, the number of photons $N$ within the setup is not fixed, but is a distribution of even photon numbers, in the range $[0,\infty)$ (in the case of purely squeezed states with no loss). The mean photon number is finite and in a following section we will discuss how to optimise experimental parameters to maximise a given photon number. Therefore in GBS we sample from photon pattern families with different total number of photons $N$,
\begin{equation}
\begin{aligned}
\{\{p_0=|\sigma_Q|\}_0&, \{p_1,p_2,...,p_{C_2}\}_2,\\
&....,\{p_1,p_2,...,p_{C_{2N}}\}_{2N},...\}\\
&=\{\{P_0\},\{P_2\},...,\{P_{2N}\},...\}
\end{aligned}
\end{equation}
with $\sum_{N=0}^\infty\{P_{2N}\}=1$. 

As with AABS we must discard events with more than one photon per mode. In addition to this, we also discard events with more photons than is allowed by the size of the interferometer and the regime we are operating in (see sections \ref{app_gbs_1} and \ref{app_gbs_2}). This means that $N<O(\sqrt{M})$ for GBS in regime 1 and $N<O(M)$ for regime 2 in order for our conditions for approximate sampling to hold.

\subsection{Photon number distribution}\label{sec:number_dist}

\begin{figure}
\includegraphics[width=.7\columnwidth, angle=90]{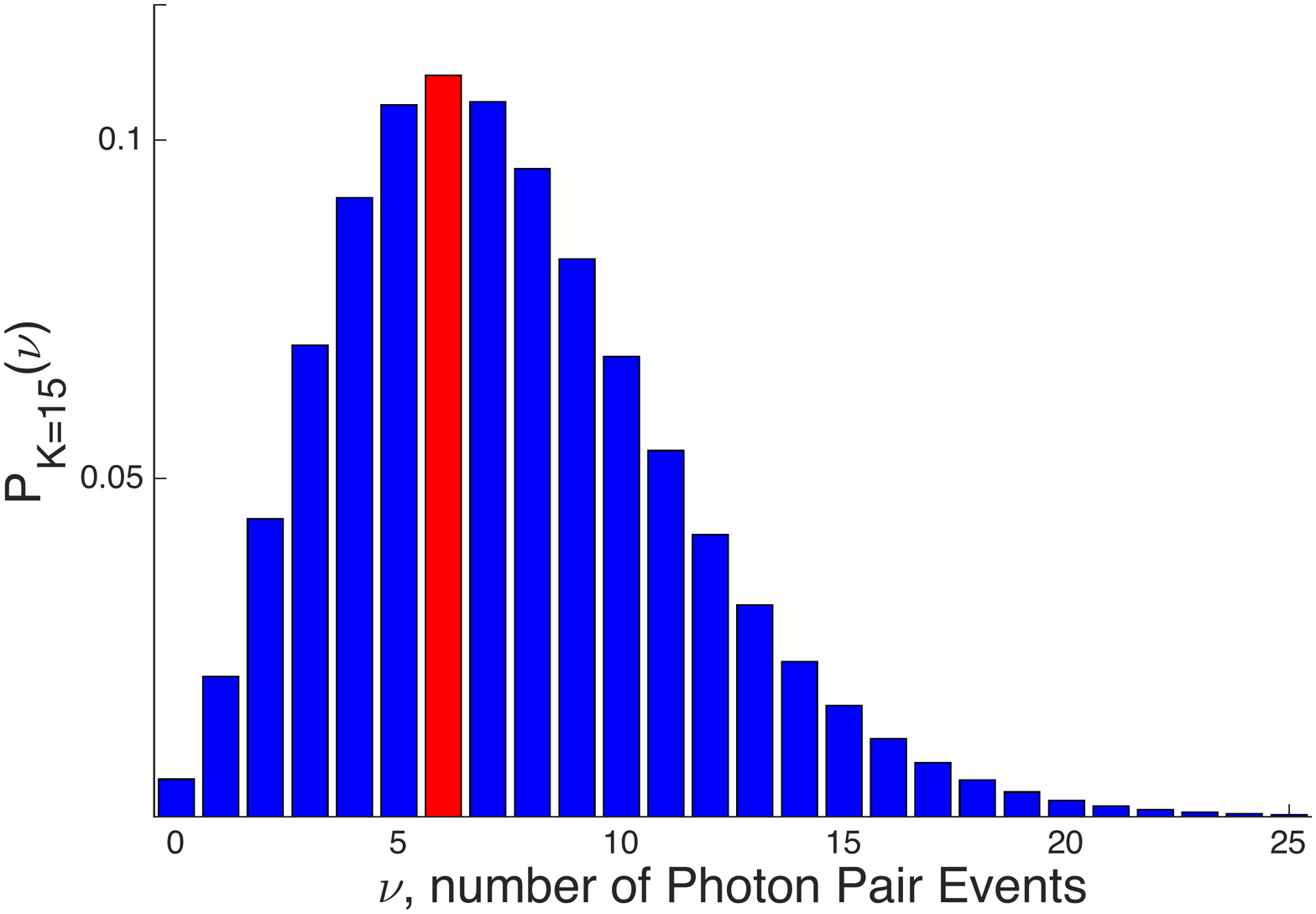}
\caption{Probabilities to generate $\nu$ photon pair events from $K=15$ single mode squeezed states with a squeezing parameter of $r=0.8814$. The modal number of this distribution is coloured red. }
\label{fig:photon-numbers}
\end{figure}

Given that squeezed states, and Gaussian states in general, produce a distribution of photon numbers and not a definitive number, we now describe that distribution and explain how to maximise the probability of the desired number of photons. We examine the probability distribution to generate $\nu$ photon pair events ($N=2\nu$ photons, as squeezed states can only produce an even number of photons). The following analysis again assumes we are working in the first regime of approximate GBS (the analysis of the second regime is more complicated). This probability to generate $2\nu$ photons from $K$ single mode squeezed states with identical squeezing parameter, from any combination of squeezers, is given by the negative binomial distribution \cite{Hilbe11}, 

\begin{equation}
\begin{aligned}
P_K(2\nu)&= {{\nu+K/2-1}\choose{\nu}} \mathrm{sech}^K(r)\mathrm{tanh}^{2\nu}(r)\\
	  &=\frac{\Gamma(\nu+K/2)}{\Gamma(K/2)\nu!}\mathrm{sech}^K(r)\mathrm{tanh}^{2\nu}(r)\, ,
\end{aligned}
\label{eq:gaussian-generation-probability}
\end{equation}
where $\Gamma(x)$ is the Gamma function. The mean number of photons is $K\sinh^2r$ and the modal number of photons (most common number) is ${n_\mathrm{modal} = 2 \lfloor (K/2-1) \sinh^2(r) \rfloor}$.
An example of this distribution is shown in Fig. \ref{fig:photon-numbers} for $K=15$ single mode squeezed states with equal squeezing parameters, $r_j=r=0.8814$. With this choice of parameter, the mean photon number per squeezer is $\langle n_\mathrm{GBS} \rangle = 1$ and the modal number, highlighted in red, is 6 photon pair events (or 12 photons). 

In an experimental setup it will be necessary to optimise the squeezing parameter to generate the desired number of photons.  This photon number is dependent on the size of the interferometer, and the number of input single mode squeezed states, which will be given by experimental resources. In principle, we can operate our GBS experiments where the number of single mode squeezed states is in the range $\nu\le K \le M$. 

If we assume that we are only interested in a specific number of photons $2\nu$, we set the squeezing parameter of all the single mode squeezed states to ensure that this is the modal number of the distribution (meaning that $2\nu$ is the most probable number of photons to be created). For $K=2\nu$ single mode squeezed states, mathematical analysis leads to $r = \ln(1+\sqrt{2})$ for large $\nu$, which means that each squeezer has a mean photon number $\langle n\rangle = 1$. If we have $K = 4\nu^2$ single mode squeezed states, which means a source at every input mode, then ${r\approx\ln(1+\sqrt{1/2\nu})}$ to set the modal number of photons to $2\nu$. 

\subsection{Computation time of hafnian relative to permanent}

The main aim of Boson sampling protocols is to generate a state that a classical computer cannot simulate in reasonable time, therefore the relative computational time of the permanent and the hafnian is important. The permanent of an $N \times N$ matrix can be calculated in $O(N2^N)$ steps, whereas the  Hafnian can be calculated in $O(2^{N/2})$ steps \cite{bjorklund2012counting}. This means that in order to achieve a comparable runtime, GBS has to sample twice the number of photons as other Boson sampling schemes. This however is not a problem, as we already obtain this factor of 2 by eliminating the heralding. This requirement also has implications for the size of the interferometer necessary, which in the worst case scenario is $4N^2$, a constant increase compared to SBS (a network of size $N^2$ is considered). 


\section{Relationship to SBS}\label{sec:GBS-vs-SBS}

In this section we now demonstrate the relationship between SBS and GBS, by describing the SBS setup in terms of GBS and can formally show the connection between the two protocols by using the relationship between the permanent and the hafnian.

\begin{figure}
\includegraphics[width=1.\columnwidth]{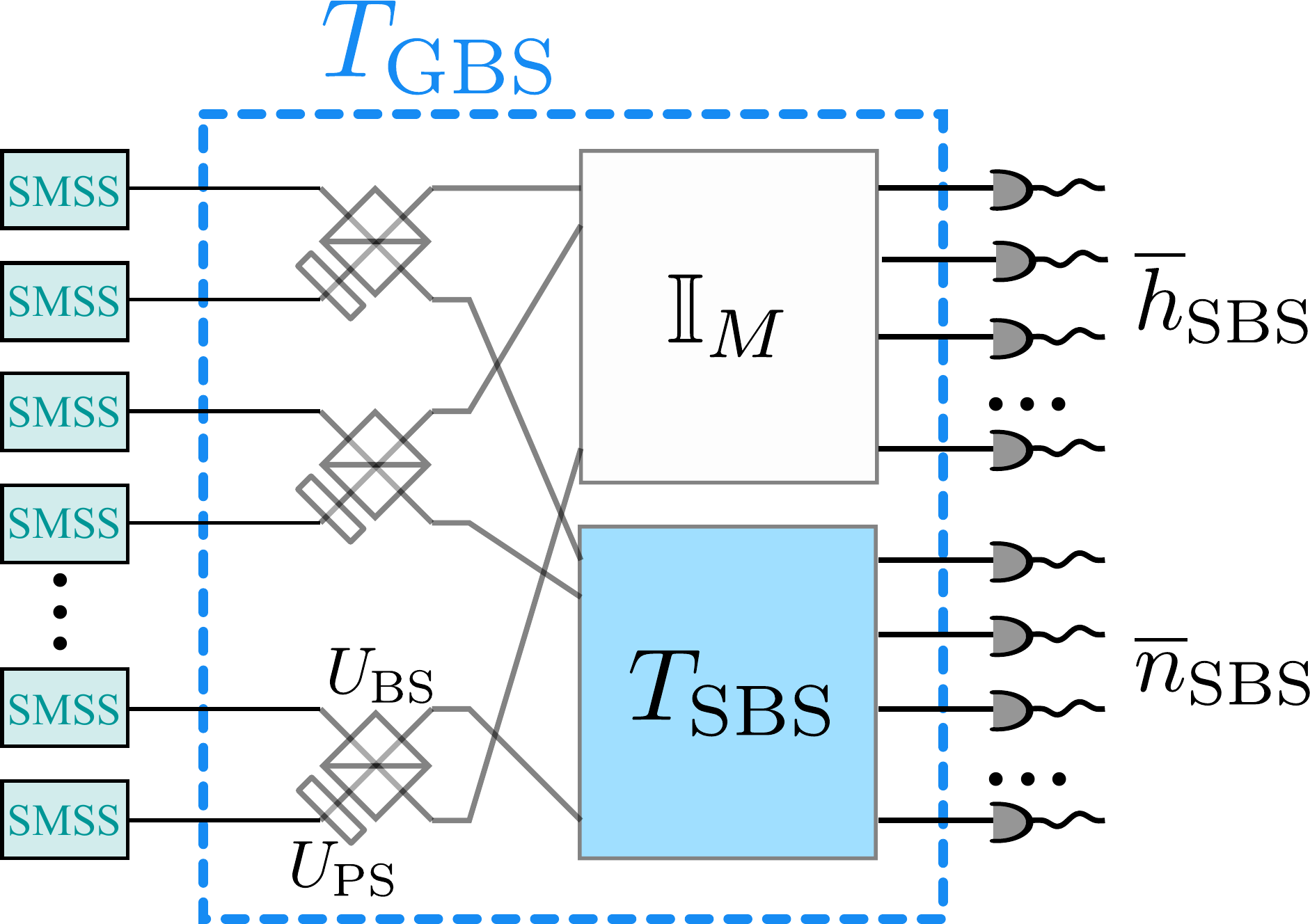}
\caption{SBS is a special case of GBS. $2M$ single mode squeezed states (SMSS) enter an array of phase shifters $U_\mathrm{PS}$ and beam splitters $U_\mathrm{BS}$ to transform them to two mode squeezed states, which are required in SBS. Then, one half of the photons is directly routed to a detection unit to generate the heralding pattern $\bar{h}_\mathrm{SBS}$, while the other half enters the interferometer $T_\mathrm{SBS}$ and generates the sampling pattern $\bar{n}_\mathrm{SBS}$. The dashed blue box enframes the corresponding GBS interferometer.}
\label{fig:SBS-GBS}
\end{figure}

Figure \ref{fig:SBS-GBS} shows a typical SBS setup. On the left of the figure, we have $2M$ (identical) single mode squeezed states, which are then combined, pairwise, at an array of phase-shifters, $U_\mathrm{PS}$, and beamsplitters, $U_\mathrm{BS}$, that are described by the two unitary transformations

\begin{equation}
U_\mathrm{PS}=\begin{pmatrix} 1 & 0\\
							  0 & i\end{pmatrix} \quad U_\mathrm{BS}=\frac{1}{\sqrt{2}}\begin{pmatrix}1 & 1\\
																							-1 & 1\end{pmatrix}\, .\vspace{0.2cm}
\end{equation}
This transformation creates the initial M two-mode squeezed states necessary for SBS. One mode of each two-mode squeezed state is sent directly to a set of detectors (i.e. transformed by the identity $\mathds{1}_M$), where the detection of a photon heralds the presence of the other photon from the photon-pair. This photon then enters the corresponding input mode of a Haar random interferometer $\mathbf T_\mathrm{SBS}$, with dimension $M$, and at the output we measure all modes to detect the position of the photons. This yields two photon patterns at the output, $\bar{n}$ for the sampled photons and $\bar{h}$ for the herald photons.  

As the input state is dependent upon the herald pattern, the probability to measure a specific pattern from an SBS experiment is actually a conditional probability, $\mathrm{Pr}(\overline{n}|\overline{h})$. We can relate this to a joint probability using Bayes' theorem. This joint probability, to measure the combined pattern $\bar{n} \cap \bar{h}$, is exactly the probability which we obtain when we consider this specific setup as a GBS experiment,
\begin{equation}
\mathrm{Pr}_\mathrm{SBS}(\overline{n})=\mathrm{Pr}(\overline{n}|\overline{h})=\frac{\mathrm{Pr}_\mathrm{GBS}(\overline{n}\cap\overline{h})}{\mathrm{Pr}(\overline{h})}\, . \label{bayes_th}
\end{equation}
The denominator in Eq. \eqref{bayes_th} is then the probability to generate the heralding pattern, which, due to the identity transformation in the herald arm, is simply the probability to generate the total number of photons that $\bar{h}$ represents. We can therefore interpret SBS as a specialised GBS experiment that samples from an interferometer of a very specific shape,
\begin{equation}
T_\mathrm{GBS}= \mathds{I}_M \oplus T_\mathrm{SBS}\times \bigoplus^M_{j=1} U_{\mathrm{BS}_j} U_{\mathrm{PS}_j}\, . \label{eq:tgbs}
\end{equation}

A more formal proof of this connection between SBS and GBS can be given by using the relationship between the permanent and the hafnian. We begin with the SBS experiment, where, for simplicity, all $M$ two-mode squeezed states have equal squeezing parameter $r$ and the generated photons then enter the interferometer $\mathds{I}_M \oplus T_\mathrm{SBS}$. The probability to measure a the sampling pattern $\bar{n}$ given a herald pattern $\bar{h}$ is
\begin{equation}
\mathrm{Pr}_\mathrm{SBS}(\overline{n}|\overline{h})=\frac{|\mathrm{Perm}(T_S)|^2}{\overline{n}!\,\overline{h}!}=\frac{\mathrm{Perm}(T_S)\mathrm{Perm}(T_S^*)}{\overline{n}!\,\overline{h}!}\, , \label{eq:pr_sbs}
\end{equation}
where $\mathbf T_S$ is the submatrix that is constructed from the input and output position of the photons. Note that the input position of the photons is given by the pattern, $\bar{h}$. To map this probability to our GBS experiments, we have to express the SBS protocol in terms of covariance matrices. The Gaussian output state after the SBS interferometer has the covariance matrix
\begin{equation}
\sigma=\frac{1}{2}(\mathds{I}\oplus T_\mathrm{SBS} \oplus \mathds{I}\oplus T_\mathrm{SBS}^*)S^{}_\mathrm{TM} S_\mathrm{TM}^\dagger(\mathds{I}\oplus T_\mathrm{SBS} \oplus\mathds{I}\oplus T_\mathrm{SBS}^*)^\dagger\, ,
\end{equation}
where 
\begin{equation}
S_\mathrm{TM}=\left (
\begin{array}{c|c}
    \cosh(r)\, \mathds{I}_{2M} & \begin{matrix} \begin{array}{c@{}c@{}} 0_M & \sinh(r) \,\mathds{I}_{M} \\ \sinh(r)\, \mathds{I}_{M} & 0_M\\  \end{array}  \end{matrix} \\
\hline
    \begin{array}{c@{}c@{}} 0_M & \sinh(r)\, \mathds{I}_{M} \\ \sinh(r)\, \mathds{I}_{M} & 0_M\\  \end{array}  & \cosh(r)\, \mathds{I}_{2M}
\end{array}
\right )
\end{equation}
which encodes the operation of the two mode squeezers (the black bars are for better clarity of the four blocks). The order of the modes is
\begin{equation}
[\hat{a}_1,...\hat{a}_M,\hat{b}_1,...,\hat{b}_M,\hat{a}^\dagger_1,...\hat{a}^\dagger_M,\hat{b}^\dagger_1,...,\hat{b}^\dagger_M],
\end{equation}
where $\hat{a}_j$ denotes the $M$ herald modes and $\hat{b}_j$ are the $M$ sampling modes. The probability for a valid GBS event in this interpretation is given by
\begin{equation}
\mathrm{Pr}_\mathrm{GBS}(\overline{n}\cap\overline{h})=\frac{\mathrm{Haf}(\mathbf{A}_S)}{\overline{n}!\,\overline{h}!\sqrt{|\sigma_Q|}}\, ,
\end{equation}
with $\sqrt{|\sigma_Q|}=\mathrm{cosh}^{2M}(r)$ for $M$ two mode squeezed states. The matrix $\mathbf{A}_S$ has a simple form and is given by
\begin{equation}
\mathbf{A}_S=-\mathrm{tanh}(r)\begin{pmatrix}0 & T_S^\dagger & 0 & 0\\
									T_S^* & 0 & 0 & 0\\
									0 & 0 & 0 & T_S^t\\
									0 & 0 & T_S & 0 \end{pmatrix} = B_S \oplus B_S^*\, .
\end{equation}
We can use Eq. \eqref{eq:permanent-hafnian-relation} to express the hafnian in terms of the permanent
\begin{equation}
\begin{aligned}
\mathrm{Haf}(\mathbf{A}_S) &= \mathrm{Haf}(B_S)\mathrm{Haf}(B_S^*)\\
				  &= \mathrm{tanh}^{2N}(r)\mathrm{Perm}(T_S)\mathrm{Perm}(T_S^*)\\
				  &= \mathrm{tanh}^{2N}(r) |\mathrm{Perm}(T_S)|^2\, .
\end{aligned}
\end{equation}
We finally arrive at
\begin{equation}
\begin{aligned}
\mathrm{Pr}_\mathrm{GBS}(\overline{n}\cap\overline{h})&
 = \frac{ \mathrm{sech}^{2M}(r)\mathrm{tanh}^{2N}(r) |\mathrm{Perm}(T_S)|^2}{\overline{n}!\, \overline{h}!} \label{eq:pr_gbs}
\end{aligned}
\end{equation}
and
\begin{equation}
\mathrm{Pr}(\overline{h})=\mathrm{sech}^{2M}(r)\, \mathrm{tanh}^{2N}(r) \label{eq:pr_hbar}
\end{equation}
Combining equations \eqref{eq:pr_gbs} and \eqref{eq:pr_hbar} and comparing to \eqref{eq:pr_sbs},  we can see that 
\begin{equation}
\frac{\mathrm{Pr}_\mathrm{GBS}(\overline{n}\cap\overline{h})}{\mathrm{Pr}(\overline{h})}=\frac{|\mathrm{Perm}(T_S)|^2}{\overline{n}!\, \overline{h}!}=\mathrm{Pr}_\mathrm{SBS}(\overline{n}|\overline{h})\, ,
\end{equation}
as expected. This demonstrates how SBS can be considered as a subset of all possible GBS experiments. 

This viewpoint also illustrates why we are allowed to retain multiple photons from the same squeezer. In GBS, we use a coherent superposition \footnote{Note that there is no phase relation  between single photons, while GBS, in contrast to AABS and SBS, relies on coherent superpositions of photon numbers and thus phase control of the input states is required.} 
over all (even) photon number states. Our ignorance of the input state in the Fock basis allows us to use `paths' where all the photons come from the same squeezer, without being able to distinguish these events from the ones where the photons come from different squeezers. Contrarily, in SBS, the herald detectors collapse our input state to a specific one, giving us exact knowledge of this state in the Fock basis.


\section{Rate of photon generation}\label{sec:efficiency}

In this section we describe one of the main advantages that GBS has in an experimental implementation, the rate of photon generation. We then compare the GBS scheme to existing Boson sampling implementations.


\subsection{Resource efficiency compared to single photon schemes}

In section \ref{sec:number_dist} we discussed the probability to generate $\nu$ photon pair events from the $K\geq 2\nu$ single mode squeezed states to saturate the complexity of the GBS scheme (Eq.~\eqref{eq:gaussian-generation-probability}). In this section we compare how this probability scales in comparison to existing Boson sampling schemes with probabilistic single photon inputs. 

PFBS protocols generate their single photon input states with a limited number of $K$ two-mode squeezers, where SBS as a special case with $N^2$ two-mode squeezers. The probability to generate $\nu$ photon pair events from $K$ two mode squeezed states and equal squeezing parameter $r$, is given by the binomial distribution \cite{Lund:2014p10967}
\begin{equation}
\mathrm{Pr}_{K,\mathrm{PFBS}}(\nu)={{K}\choose{\nu}}\mathrm{sech}^{2K}(r)\mathrm{tanh}^{2\nu}(r)\, .
\end{equation}
The ratio of this and Eq.~\eqref{eq:gaussian-generation-probability} to generate $\nu$ photon pairs from $K$ two mode squeezed states for PFBS, and $2K$ single mode squeezed states for GBS (as a fair comparison) is (for the same squeezing parameter $r$) 
\begin{equation}
\begin{aligned}
\frac{\mathrm{Pr}_\mathrm{PFBS}(\nu)}{\mathrm{Pr}_\mathrm{GBS}(\nu)}&={{K}\choose{\nu}}\left[{{K+\nu-1}\choose{\nu}}\right]^{-1}\\
			&= \frac{K!(K-1)!}{(K-\nu)!(K+\nu-1)!}\\
			&=\lim_{N\rightarrow \infty, K>\nu} \left(\frac{K-\nu}{K-1}\right)^\nu\, .
\end{aligned}
\end{equation}
This ratio scales exponentially in favour of GBS, with an improvement of roughly $\nu^\nu$. We can explain this behaviour by the all the possible ways to generate $\nu$ photon pairs in total in each protocol. While PFBS is restricted to a single photon pair event per squeezer, GBS is not hindered by this restriction and can use multiple photon pairs from the same squeezers, signified by the extra term $(\nu-1)$ in the binomial factor. In the special case of SBS with $N^2$ squeezers, this number converges to Euler's number $e$.

However, we also note that in GBS we do not have to implement $\nu^2$ squeezers at the input to saturate the complexity of the sampling problem, but only $2\nu$. Therefore, compared to SBS, we can save a quadratic factor in the number of squeezers.


\subsection{Comparison to current sources}

\begin{figure}
\includegraphics[width=.9\columnwidth]{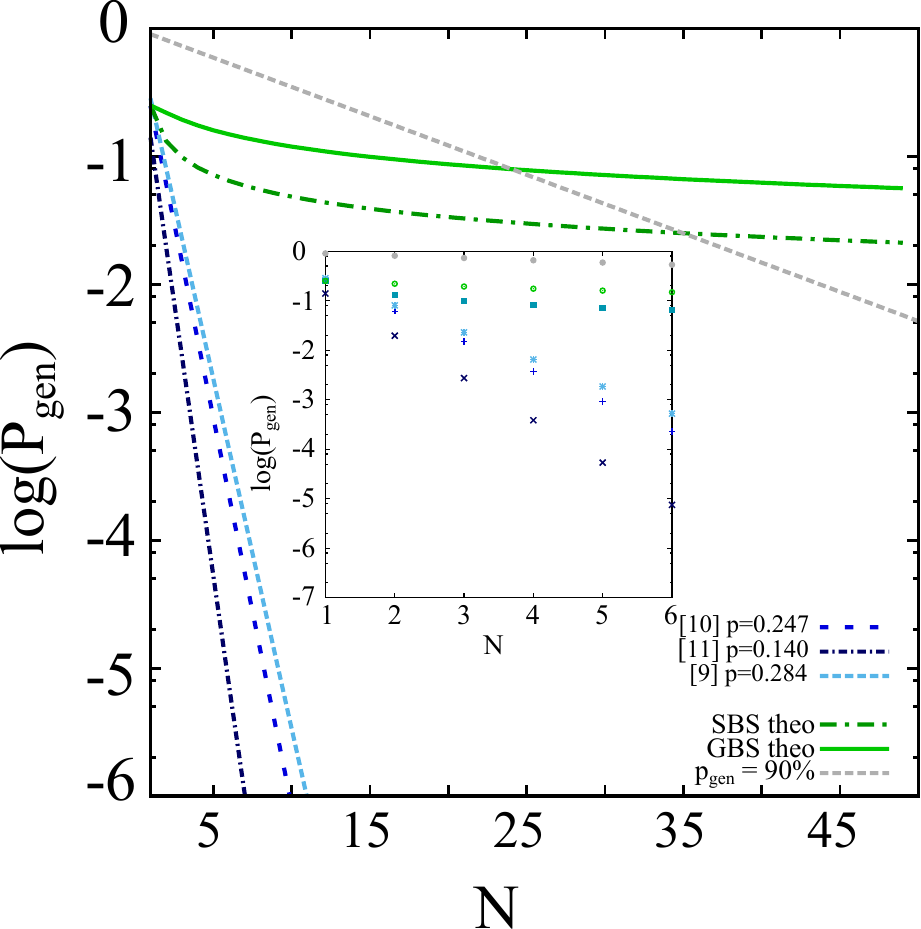}
\caption{Comparison of single photon efficiency for different Boson Sampling approaches. The first three lines represent the current state of the art with solid state sources \cite{he_scalable_2016, loredo_boson_2017, wang2017high}. In comparison, we plot the scaling performance of SBS \cite{Lund:2014p10967} and GBS and an almost optimal deterministic source with 90\% efficiency. Even for this high value, GBS ($K=N^2$) is advantageous for more than $\approx\,25$ photons and SBS for more than $\approx 35$ photons.}
\label{fig:compare-sources}
\end{figure}

To compare the GBS approach with existing protocols, we plot the probabilities to obtain $N$ photons from different types of sources in figure \ref{fig:compare-sources}. We first compare the single photon efficiency $p=p_{\small gen}p_{\small extr}$, which we define as the product of the generation probability,  $p_{\mbox{\small gen}}$, and the extraction probability, $p_{\mbox{\small extr}}$,  of state-of-the-art solid state sources from He et al. \cite{he_scalable_2016} (dashed blue line, $p=0.247$), Loredo et al. \cite{loredo_boson_2017} (blue dash-dotted line, $p=0.14$) and Wang et al. \cite{wang2017high} (densely dashed blue line, $p=0.284$) where for the latter we use the efficiency of the demultiplexer implemented to inject photons in different inputs of the boson sampler ($p_{dem}$=0.845) as an additional factor for the single photon efficiency $p=0.337 p_{dem}$.
All of these approaches converge exponentially to zero for high $N$ and only differ in their single photon success probability. The green dash-dotted line shows the theoretical SBS scaling to higher photon numbers (proportional to $\tfrac{1}{\sqrt{N}}$)\footnote{The experimental implementation of \cite{bentivegna2015experimental} does not use as many photon pair sources as the number of modes (9- and 13- mode unitaries with $K=6$ photon pair sources), for this reason we do not report a scaling of their approach}. 

Finally, we plot the theoretical scaling of our GBS protocol for $K=N^2$ sources with the green, solid line. We observe the $e$-fold improvement towards the SBS schemes and the expected $\tfrac{1}{\sqrt{N}}$ scaling. For comparison, we also show the scaling behaviour of an almost perfect single photon source with $90\%$ generation probability (grey dashed line). Even in this case, the polynomial scaling of the Gaussian protocols allows for better generation probabilities in the high photon number regime; the break-even point for GBS is around 25 photons, while the one for SBS is higher with 35 photons. As the "interesting" regime for Boson Sampling experiments begins around $N=50-100$ photons \cite{aaronson_computational_2011, neville2017classical, clifford2017}, Gaussian protocols are more likely to reach the required photon numbers with reasonable generation rates. Indeed, this break-even point can already be reached with existing sources of parametric down-conversion \cite{harder_optimized_2013, harder_single-mode_2016}.


\section{Conclusions}\label{sec:conclusion}

In this paper, we have demonstrated how to use the full nature of squeezed states to construct a Boson sampling protocol and extended our results and analysis from \cite{hamilton_gaussian_2016}. First, we derived a new expression for the probability to measure a specific photon sampling pattern from a general Gaussian state, which depends upon the hafnian, a matrix function more general than the permanent. Our work in this paper extends this formula to include displacements contributions, so that all Gaussian states are covered, and we also discussed how to include higher order detection events into our formalism.
Following this we discuss a Boson sampling protocol, using squeezed states entering a linear interferometer, which is based on the fact that to calculate the hafnian is a \#P problem. We then propose arguments why approximate sampling from Gaussian states is also a \#P problem and explained the various requirements for the complexity in GBS to be satisfied. 
Furthermore, we related our protocol to the most general protocol up to date SBS, and showed that it is only a restricted subset of our GBS scheme. Finally, we compared the theoretical generation probability of GBS with the actual generation rates of current experiments, showing the promise of sampling squeezed states instead of single photons. 

Within experimental quantum optics, starting with a squeezed state, using linear optical transformations and postselecting measurement outcomes is a very common method to create different families of photonic states, and is universal for quantum computation. We can model this situation with GBS if we ``move'' all the measurements to the end of the computation, after the linear optical elements. This means that the GBS protocol includes other photonic Boson sampling protocols as special cases, which we have demonstrated here with SBS, but also
those problems involving Schr\"{o}dinger cat states and photon added /subtracted states \cite{Rohde:2015p11547, Olson:2015p10963, Seshadreesan:2015p12695}. We also note that due to the time-reversal symmetry of quantum mechanics, GBS also includes the situation of Fock Boson sampling with Gaussian measurements \cite{chakhmakhchyan2017, lund_exact_2017, chabaud2017}.  

Another important aspect in Boson sampling schemes is the verification of the correct operation of the device in an efficient manner \cite{gogolin2013boson, aaronson2013bosonsampling, Carolan2014, Spagnolo:2014p10480,bentivegna2014bayesian}. As the size of the output state space with single photons is exponentially large, full state tomography would be a practically impossible task. In recent works \cite{Tichy2014, walschaers2016statistical}, statistical averages that can be calculated were used to verify the device operation. As Gaussian states are completely characterised by their covariance matrix, which is of size $M^2$ and can be efficiently measured \cite{Rehacek2009}, then an interesting question is if this information can be used, in combination with the methods developed in the continuous variable field, to help verify the correct operation of the device. 

While Boson sampling is demanding and makes use of experiments at their full capabilities, we show here, through GBS, a new regime with advantages that will bring the protocol within the reach of current technology. 

\textbf{Acknowledgements: }
This work has received funding from the European Union's Horizon 2020 research and innovation program under the QUCHIP project Grant No.~641039. C.S.H. and I. J. received support from the Grant Agency of the Czech Republic under grant No. GACR 17-00844S and the Ministry of Education RVO 68407700.

The authors would like to thank A. Arkhipov, T. C. Ralph, A. Bj\"orklund, S. Rahimi-Keshari and T. Weich for useful comments. 

\bibliography{gbs_pra}

\begin{thebibliography}{57}%
\makeatletter
\providecommand \@ifxundefined [1]{%
 \@ifx{#1\undefined}
}%
\providecommand \@ifnum [1]{%
 \ifnum #1\expandafter \@firstoftwo
 \else \expandafter \@secondoftwo
 \fi
}%
\providecommand \@ifx [1]{%
 \ifx #1\expandafter \@firstoftwo
 \else \expandafter \@secondoftwo
 \fi
}%
\providecommand \natexlab [1]{#1}%
\providecommand \enquote  [1]{``#1''}%
\providecommand \bibnamefont  [1]{#1}%
\providecommand \bibfnamefont [1]{#1}%
\providecommand \citenamefont [1]{#1}%
\providecommand \href@noop [0]{\@secondoftwo}%
\providecommand \href [0]{\begingroup \@sanitize@url \@href}%
\providecommand \@href[1]{\@@startlink{#1}\@@href}%
\providecommand \@@href[1]{\endgroup#1\@@endlink}%
\providecommand \@sanitize@url [0]{\catcode `\\12\catcode `\$12\catcode
  `\&12\catcode `\#12\catcode `\^12\catcode `\_12\catcode `\%12\relax}%
\providecommand \@@startlink[1]{}%
\providecommand \@@endlink[0]{}%
\providecommand \url  [0]{\begingroup\@sanitize@url \@url }%
\providecommand \@url [1]{\endgroup\@href {#1}{\urlprefix }}%
\providecommand \urlprefix  [0]{URL }%
\providecommand \Eprint [0]{\href }%
\providecommand \doibase [0]{http://dx.doi.org/}%
\providecommand \selectlanguage [0]{\@gobble}%
\providecommand \bibinfo  [0]{\@secondoftwo}%
\providecommand \bibfield  [0]{\@secondoftwo}%
\providecommand \translation [1]{[#1]}%
\providecommand \BibitemOpen [0]{}%
\providecommand \bibitemStop [0]{}%
\providecommand \bibitemNoStop [0]{.\EOS\space}%
\providecommand \EOS [0]{\spacefactor3000\relax}%
\providecommand \BibitemShut  [1]{\csname bibitem#1\endcsname}%
\let\auto@bib@innerbib\@empty
\bibitem [{\citenamefont {Aaronson}\ and\ \citenamefont
  {Arkhipov}(2011)}]{aaronson_computational_2011}%
  \BibitemOpen
  \bibfield  {author} {\bibinfo {author} {\bibfnamefont {S.}~\bibnamefont
  {Aaronson}}\ and\ \bibinfo {author} {\bibfnamefont {A.}~\bibnamefont
  {Arkhipov}},\ }in\ \href {\doibase 10.1145/1993636.1993682} {\emph {\bibinfo
  {booktitle} {Proceedings of the forty-third annual ACM symposium on Theory of
  computing}}}\ (\bibinfo {organization} {ACM},\ \bibinfo {year} {2011})\ pp.\
  \bibinfo {pages} {333--342}\BibitemShut {NoStop}%
\bibitem [{\citenamefont {Aaronson}\ and\ \citenamefont
  {Arkhipov}(2013{\natexlab{a}})}]{Aaronson:2013p7598}%
  \BibitemOpen
  \bibfield  {author} {\bibinfo {author} {\bibfnamefont {S.}~\bibnamefont
  {Aaronson}}\ and\ \bibinfo {author} {\bibfnamefont {A.}~\bibnamefont
  {Arkhipov}},\ }\href {\doibase 10.4086/toc.2013.v009a004} {\bibfield
  {journal} {\bibinfo  {journal} {Theory of Comput.}\ }\textbf {\bibinfo
  {volume} {9}},\ \bibinfo {pages} {143} (\bibinfo {year}
  {2013}{\natexlab{a}})}\BibitemShut {NoStop}%
\bibitem [{\citenamefont {Broome}\ \emph {et~al.}(2013)\citenamefont {Broome},
  \citenamefont {Fedrizzi}, \citenamefont {Rahimi-Keshari}, \citenamefont
  {Dove}, \citenamefont {Aaronson}, \citenamefont {Ralph},\ and\ \citenamefont
  {White}}]{Broome:2013p7136}%
  \BibitemOpen
  \bibfield  {author} {\bibinfo {author} {\bibfnamefont {M.~A.}\ \bibnamefont
  {Broome}}, \bibinfo {author} {\bibfnamefont {A.}~\bibnamefont {Fedrizzi}},
  \bibinfo {author} {\bibfnamefont {S.}~\bibnamefont {Rahimi-Keshari}},
  \bibinfo {author} {\bibfnamefont {J.}~\bibnamefont {Dove}}, \bibinfo {author}
  {\bibfnamefont {S.}~\bibnamefont {Aaronson}}, \bibinfo {author}
  {\bibfnamefont {T.~C.}\ \bibnamefont {Ralph}}, \ and\ \bibinfo {author}
  {\bibfnamefont {A.~G.}\ \bibnamefont {White}},\ }\href {\doibase
  10.1126/science.1231440} {\bibfield  {journal} {\bibinfo  {journal}
  {Science}\ }\textbf {\bibinfo {volume} {339}},\ \bibinfo {pages} {794}
  (\bibinfo {year} {2013})}\BibitemShut {NoStop}%
\bibitem [{\citenamefont {Tillmann}\ \emph {et~al.}(2013)\citenamefont
  {Tillmann}, \citenamefont {Daki{\'c}}, \citenamefont {Heilmann},
  \citenamefont {Nolte}, \citenamefont {Szameit},\ and\ \citenamefont
  {Walther}}]{Tillmann:2013p10461}%
  \BibitemOpen
  \bibfield  {author} {\bibinfo {author} {\bibfnamefont {M.}~\bibnamefont
  {Tillmann}}, \bibinfo {author} {\bibfnamefont {B.}~\bibnamefont {Daki{\'c}}},
  \bibinfo {author} {\bibfnamefont {R.}~\bibnamefont {Heilmann}}, \bibinfo
  {author} {\bibfnamefont {S.}~\bibnamefont {Nolte}}, \bibinfo {author}
  {\bibfnamefont {A.}~\bibnamefont {Szameit}}, \ and\ \bibinfo {author}
  {\bibfnamefont {P.}~\bibnamefont {Walther}},\ }\href {\doibase
  10.1038/nphoton.2013.102} {\bibfield  {journal} {\bibinfo  {journal} {Nature
  Photonics}\ }\textbf {\bibinfo {volume} {7}},\ \bibinfo {pages} {540}
  (\bibinfo {year} {2013})}\BibitemShut {NoStop}%
\bibitem [{\citenamefont {Spring}\ \emph {et~al.}(2013)\citenamefont {Spring},
  \citenamefont {Metcalf}, \citenamefont {Humphreys}, \citenamefont
  {Kolthammer}, \citenamefont {Jin}, \citenamefont {Barbieri}, \citenamefont
  {Datta}, \citenamefont {Thomas-Peter}, \citenamefont {Langford},
  \citenamefont {Kundys}, \citenamefont {Gates}, \citenamefont {Smith},
  \citenamefont {Smith},\ and\ \citenamefont {Walmsley}}]{Spring:2013p7137}%
  \BibitemOpen
  \bibfield  {author} {\bibinfo {author} {\bibfnamefont {J.~B.}\ \bibnamefont
  {Spring}}, \bibinfo {author} {\bibfnamefont {B.~J.}\ \bibnamefont {Metcalf}},
  \bibinfo {author} {\bibfnamefont {P.~C.}\ \bibnamefont {Humphreys}}, \bibinfo
  {author} {\bibfnamefont {W.~S.}\ \bibnamefont {Kolthammer}}, \bibinfo
  {author} {\bibfnamefont {X.-M.}\ \bibnamefont {Jin}}, \bibinfo {author}
  {\bibfnamefont {M.}~\bibnamefont {Barbieri}}, \bibinfo {author}
  {\bibfnamefont {A.}~\bibnamefont {Datta}}, \bibinfo {author} {\bibfnamefont
  {N.}~\bibnamefont {Thomas-Peter}}, \bibinfo {author} {\bibfnamefont {N.~K.}\
  \bibnamefont {Langford}}, \bibinfo {author} {\bibfnamefont {D.}~\bibnamefont
  {Kundys}}, \bibinfo {author} {\bibfnamefont {J.~C.}\ \bibnamefont {Gates}},
  \bibinfo {author} {\bibfnamefont {B.~J.}\ \bibnamefont {Smith}}, \bibinfo
  {author} {\bibfnamefont {P.~G.~R.}\ \bibnamefont {Smith}}, \ and\ \bibinfo
  {author} {\bibfnamefont {I.~A.}\ \bibnamefont {Walmsley}},\ }\href {\doibase
  10.1126/science.1231692} {\bibfield  {journal} {\bibinfo  {journal}
  {Science}\ }\textbf {\bibinfo {volume} {339}},\ \bibinfo {pages} {798}
  (\bibinfo {year} {2013})}\BibitemShut {NoStop}%
\bibitem [{\citenamefont {Crespi}\ \emph {et~al.}(2013)\citenamefont {Crespi},
  \citenamefont {Osellame}, \citenamefont {Ramponi}, \citenamefont {Brod},
  \citenamefont {ao}, \citenamefont {Spagnolo}, \citenamefont {Vitelli},
  \citenamefont {Maiorino}, \citenamefont {Mataloni},\ and\ \citenamefont
  {Sciarrino}}]{cres13npo2}%
  \BibitemOpen
  \bibfield  {author} {\bibinfo {author} {\bibfnamefont {A.}~\bibnamefont
  {Crespi}}, \bibinfo {author} {\bibfnamefont {R.}~\bibnamefont {Osellame}},
  \bibinfo {author} {\bibfnamefont {R.}~\bibnamefont {Ramponi}}, \bibinfo
  {author} {\bibfnamefont {D.~J.}\ \bibnamefont {Brod}}, \bibinfo {author}
  {\bibfnamefont {E.~F.~G.}\ \bibnamefont {ao}}, \bibinfo {author}
  {\bibfnamefont {N.}~\bibnamefont {Spagnolo}}, \bibinfo {author}
  {\bibfnamefont {C.}~\bibnamefont {Vitelli}}, \bibinfo {author} {\bibfnamefont
  {E.}~\bibnamefont {Maiorino}}, \bibinfo {author} {\bibfnamefont
  {P.}~\bibnamefont {Mataloni}}, \ and\ \bibinfo {author} {\bibfnamefont
  {F.}~\bibnamefont {Sciarrino}},\ }\href@noop {} {\bibfield  {journal}
  {\bibinfo  {journal} {Nature Photonics}\ }\textbf {\bibinfo {volume} {7}},\
  \bibinfo {pages} {545} (\bibinfo {year} {2013})}\BibitemShut {NoStop}%
\bibitem [{\citenamefont {Neville}\ \emph {et~al.}(2017)\citenamefont
  {Neville}, \citenamefont {Sparrow}, \citenamefont {Clifford}, \citenamefont
  {Johnston}, \citenamefont {Birchall}, \citenamefont {Montanaro},\ and\
  \citenamefont {Laing}}]{neville2017classical}%
  \BibitemOpen
  \bibfield  {author} {\bibinfo {author} {\bibfnamefont {A.}~\bibnamefont
  {Neville}}, \bibinfo {author} {\bibfnamefont {C.}~\bibnamefont {Sparrow}},
  \bibinfo {author} {\bibfnamefont {R.}~\bibnamefont {Clifford}}, \bibinfo
  {author} {\bibfnamefont {E.}~\bibnamefont {Johnston}}, \bibinfo {author}
  {\bibfnamefont {P.~M.}\ \bibnamefont {Birchall}}, \bibinfo {author}
  {\bibfnamefont {A.}~\bibnamefont {Montanaro}}, \ and\ \bibinfo {author}
  {\bibfnamefont {A.}~\bibnamefont {Laing}},\ }\href {\doibase
  10.1038/nphys4270} {\bibfield  {journal} {\bibinfo  {journal} {Nature
  Physics}\ }\textbf {\bibinfo {volume} {13}},\ \bibinfo {pages} {1153}
  (\bibinfo {year} {2017})}\BibitemShut {NoStop}%
\bibitem [{\citenamefont {Clifford}\ and\ \citenamefont
  {Clifford}(2017)}]{clifford2017}%
  \BibitemOpen
  \bibfield  {author} {\bibinfo {author} {\bibfnamefont {P.}~\bibnamefont
  {Clifford}}\ and\ \bibinfo {author} {\bibfnamefont {R.}~\bibnamefont
  {Clifford}},\ }\href {http://arxiv.org/abs/1706.01260} {\bibfield  {journal}
  {\bibinfo  {journal} {CoRR}\ }\textbf {\bibinfo {volume} {abs/1706.01260}}
  (\bibinfo {year} {2017})},\ \Eprint {http://arxiv.org/abs/1706.01260}
  {arXiv:1706.01260} \BibitemShut {NoStop}%
\bibitem [{\citenamefont {Wang}\ \emph {et~al.}(2017)\citenamefont {Wang},
  \citenamefont {He}, \citenamefont {Li}, \citenamefont {Su}, \citenamefont
  {Li}, \citenamefont {Huang}, \citenamefont {Ding}, \citenamefont {Chen},
  \citenamefont {Liu}, \citenamefont {Qin} \emph {et~al.}}]{wang2017high}%
  \BibitemOpen
  \bibfield  {author} {\bibinfo {author} {\bibfnamefont {H.}~\bibnamefont
  {Wang}}, \bibinfo {author} {\bibfnamefont {Y.}~\bibnamefont {He}}, \bibinfo
  {author} {\bibfnamefont {Y.-H.}\ \bibnamefont {Li}}, \bibinfo {author}
  {\bibfnamefont {Z.-E.}\ \bibnamefont {Su}}, \bibinfo {author} {\bibfnamefont
  {B.}~\bibnamefont {Li}}, \bibinfo {author} {\bibfnamefont {H.-L.}\
  \bibnamefont {Huang}}, \bibinfo {author} {\bibfnamefont {X.}~\bibnamefont
  {Ding}}, \bibinfo {author} {\bibfnamefont {M.-C.}\ \bibnamefont {Chen}},
  \bibinfo {author} {\bibfnamefont {C.}~\bibnamefont {Liu}}, \bibinfo {author}
  {\bibfnamefont {J.}~\bibnamefont {Qin}},  \emph {et~al.},\ }\href@noop {}
  {\bibfield  {journal} {\bibinfo  {journal} {Nature Photonics}\ }\textbf
  {\bibinfo {volume} {11}},\ \bibinfo {pages} {361} (\bibinfo {year}
  {2017})}\BibitemShut {NoStop}%
\bibitem [{\citenamefont {He}\ \emph {et~al.}(2017)\citenamefont {He},
  \citenamefont {Ding}, \citenamefont {Su}, \citenamefont {Huang},
  \citenamefont {Qin}, \citenamefont {Wang}, \citenamefont {Unsleber},
  \citenamefont {Chen}, \citenamefont {Wang}, \citenamefont {He}, \citenamefont
  {Wang}, \citenamefont {Zhang}, \citenamefont {Chen}, \citenamefont
  {Schneider}, \citenamefont {Kamp}, \citenamefont {You}, \citenamefont {Wang},
  \citenamefont {H\"ofling}, \citenamefont {Lu},\ and\ \citenamefont
  {Pan}}]{he_scalable_2016}%
  \BibitemOpen
  \bibfield  {author} {\bibinfo {author} {\bibfnamefont {Y.}~\bibnamefont
  {He}}, \bibinfo {author} {\bibfnamefont {X.}~\bibnamefont {Ding}}, \bibinfo
  {author} {\bibfnamefont {Z.-E.}\ \bibnamefont {Su}}, \bibinfo {author}
  {\bibfnamefont {H.-L.}\ \bibnamefont {Huang}}, \bibinfo {author}
  {\bibfnamefont {J.}~\bibnamefont {Qin}}, \bibinfo {author} {\bibfnamefont
  {C.}~\bibnamefont {Wang}}, \bibinfo {author} {\bibfnamefont {S.}~\bibnamefont
  {Unsleber}}, \bibinfo {author} {\bibfnamefont {C.}~\bibnamefont {Chen}},
  \bibinfo {author} {\bibfnamefont {H.}~\bibnamefont {Wang}}, \bibinfo {author}
  {\bibfnamefont {Y.-M.}\ \bibnamefont {He}}, \bibinfo {author} {\bibfnamefont
  {X.-L.}\ \bibnamefont {Wang}}, \bibinfo {author} {\bibfnamefont {W.-J.}\
  \bibnamefont {Zhang}}, \bibinfo {author} {\bibfnamefont {S.-J.}\ \bibnamefont
  {Chen}}, \bibinfo {author} {\bibfnamefont {C.}~\bibnamefont {Schneider}},
  \bibinfo {author} {\bibfnamefont {M.}~\bibnamefont {Kamp}}, \bibinfo {author}
  {\bibfnamefont {L.-X.}\ \bibnamefont {You}}, \bibinfo {author} {\bibfnamefont
  {Z.}~\bibnamefont {Wang}}, \bibinfo {author} {\bibfnamefont {S.}~\bibnamefont
  {H\"ofling}}, \bibinfo {author} {\bibfnamefont {C.-Y.}\ \bibnamefont {Lu}}, \
  and\ \bibinfo {author} {\bibfnamefont {J.-W.}\ \bibnamefont {Pan}},\ }\href
  {\doibase 10.1103/PhysRevLett.118.190501} {\bibfield  {journal} {\bibinfo
  {journal} {Physical Review Letters}\ }\textbf {\bibinfo {volume} {118}},\
  \bibinfo {pages} {190501} (\bibinfo {year} {2017})}\BibitemShut {NoStop}%
\bibitem [{\citenamefont {Loredo}\ \emph {et~al.}(2017)\citenamefont {Loredo},
  \citenamefont {Broome}, \citenamefont {Hilaire}, \citenamefont {Gazzano},
  \citenamefont {Sagnes}, \citenamefont {Lemaitre}, \citenamefont {Almeida},
  \citenamefont {Senellart},\ and\ \citenamefont {White}}]{loredo_boson_2017}%
  \BibitemOpen
  \bibfield  {author} {\bibinfo {author} {\bibfnamefont {J.}~\bibnamefont
  {Loredo}}, \bibinfo {author} {\bibfnamefont {M.}~\bibnamefont {Broome}},
  \bibinfo {author} {\bibfnamefont {P.}~\bibnamefont {Hilaire}}, \bibinfo
  {author} {\bibfnamefont {O.}~\bibnamefont {Gazzano}}, \bibinfo {author}
  {\bibfnamefont {I.}~\bibnamefont {Sagnes}}, \bibinfo {author} {\bibfnamefont
  {A.}~\bibnamefont {Lemaitre}}, \bibinfo {author} {\bibfnamefont
  {M.}~\bibnamefont {Almeida}}, \bibinfo {author} {\bibfnamefont
  {P.}~\bibnamefont {Senellart}}, \ and\ \bibinfo {author} {\bibfnamefont
  {A.}~\bibnamefont {White}},\ }\href {\doibase 10.1103/PhysRevLett.118.130503}
  {\bibfield  {journal} {\bibinfo  {journal} {Physical Review Letters}\
  }\textbf {\bibinfo {volume} {118}},\ \bibinfo {pages} {130503} (\bibinfo
  {year} {2017})}\BibitemShut {NoStop}%
\bibitem [{\citenamefont {Lund}\ \emph {et~al.}(2014)\citenamefont {Lund},
  \citenamefont {Laing}, \citenamefont {Rahimi-Keshari}, \citenamefont
  {Rudolph}, \citenamefont {O'Brien},\ and\ \citenamefont
  {Ralph}}]{Lund:2014p10967}%
  \BibitemOpen
  \bibfield  {author} {\bibinfo {author} {\bibfnamefont {A.~P.}\ \bibnamefont
  {Lund}}, \bibinfo {author} {\bibfnamefont {A.}~\bibnamefont {Laing}},
  \bibinfo {author} {\bibfnamefont {S.}~\bibnamefont {Rahimi-Keshari}},
  \bibinfo {author} {\bibfnamefont {T.}~\bibnamefont {Rudolph}}, \bibinfo
  {author} {\bibfnamefont {J.~L.}\ \bibnamefont {O'Brien}}, \ and\ \bibinfo
  {author} {\bibfnamefont {T.~C.}\ \bibnamefont {Ralph}},\ }\href {\doibase
  10.1103/PhysRevLett.113.100502} {\bibfield  {journal} {\bibinfo  {journal}
  {Physical Review Letters}\ }\textbf {\bibinfo {volume} {113}},\ \bibinfo
  {pages} {100502} (\bibinfo {year} {2014})}\BibitemShut {NoStop}%
\bibitem [{\citenamefont {Barkhofen}\ \emph {et~al.}(2017)\citenamefont
  {Barkhofen}, \citenamefont {Bartley}, \citenamefont {Sansoni}, \citenamefont
  {Kruse}, \citenamefont {Hamilton}, \citenamefont {Jex},\ and\ \citenamefont
  {Silberhorn}}]{Barkhofen:2017p13761}%
  \BibitemOpen
  \bibfield  {author} {\bibinfo {author} {\bibfnamefont {S.}~\bibnamefont
  {Barkhofen}}, \bibinfo {author} {\bibfnamefont {T.~J.}\ \bibnamefont
  {Bartley}}, \bibinfo {author} {\bibfnamefont {L.}~\bibnamefont {Sansoni}},
  \bibinfo {author} {\bibfnamefont {R.}~\bibnamefont {Kruse}}, \bibinfo
  {author} {\bibfnamefont {C.~S.}\ \bibnamefont {Hamilton}}, \bibinfo {author}
  {\bibfnamefont {I.}~\bibnamefont {Jex}}, \ and\ \bibinfo {author}
  {\bibfnamefont {C.}~\bibnamefont {Silberhorn}},\ }\href {\doibase
  10.1103/PhysRevLett.118.020502} {\bibfield  {journal} {\bibinfo  {journal}
  {Physical Review Letters}\ }\textbf {\bibinfo {volume} {118}},\ \bibinfo
  {pages} {020502} (\bibinfo {year} {2017})}\BibitemShut {NoStop}%
\bibitem [{\citenamefont {Yoshikawa}\ \emph {et~al.}(2016)\citenamefont
  {Yoshikawa}, \citenamefont {Yokoyama}, \citenamefont {Kaji}, \citenamefont
  {Sornphiphatphong}, \citenamefont {Shiozawa}, \citenamefont {Makino},\ and\
  \citenamefont {Furusawa}}]{yoshikawa2016}%
  \BibitemOpen
  \bibfield  {author} {\bibinfo {author} {\bibfnamefont {J.-i.}\ \bibnamefont
  {Yoshikawa}}, \bibinfo {author} {\bibfnamefont {S.}~\bibnamefont {Yokoyama}},
  \bibinfo {author} {\bibfnamefont {T.}~\bibnamefont {Kaji}}, \bibinfo {author}
  {\bibfnamefont {C.}~\bibnamefont {Sornphiphatphong}}, \bibinfo {author}
  {\bibfnamefont {Y.}~\bibnamefont {Shiozawa}}, \bibinfo {author}
  {\bibfnamefont {K.}~\bibnamefont {Makino}}, \ and\ \bibinfo {author}
  {\bibfnamefont {A.}~\bibnamefont {Furusawa}},\ }\href {\doibase
  10.1063/1.4962732} {\bibfield  {journal} {\bibinfo  {journal} {APL
  Photonics}\ }\textbf {\bibinfo {volume} {1}},\ \bibinfo {pages} {060801}
  (\bibinfo {year} {2016})}\BibitemShut {NoStop}%
\bibitem [{\citenamefont {Rahimi-Keshari}\ \emph {et~al.}(2015)\citenamefont
  {Rahimi-Keshari}, \citenamefont {Lund},\ and\ \citenamefont
  {Ralph}}]{RahimiKeshari:2015p11006}%
  \BibitemOpen
  \bibfield  {author} {\bibinfo {author} {\bibfnamefont {S.}~\bibnamefont
  {Rahimi-Keshari}}, \bibinfo {author} {\bibfnamefont {A.~P.}\ \bibnamefont
  {Lund}}, \ and\ \bibinfo {author} {\bibfnamefont {T.~C.}\ \bibnamefont
  {Ralph}},\ }\href {\doibase 10.1103/PhysRevLett.114.060501} {\bibfield
  {journal} {\bibinfo  {journal} {Physical Review Letters}\ }\textbf {\bibinfo
  {volume} {114}},\ \bibinfo {pages} {060501} (\bibinfo {year}
  {2015})}\BibitemShut {NoStop}%
\bibitem [{\citenamefont {Hamilton}\ \emph {et~al.}(2017)\citenamefont
  {Hamilton}, \citenamefont {Kruse}, \citenamefont {Sansoni}, \citenamefont
  {Barkhofen}, \citenamefont {Silberhorn},\ and\ \citenamefont
  {Jex}}]{hamilton_gaussian_2016}%
  \BibitemOpen
  \bibfield  {author} {\bibinfo {author} {\bibfnamefont {C.~S.}\ \bibnamefont
  {Hamilton}}, \bibinfo {author} {\bibfnamefont {R.}~\bibnamefont {Kruse}},
  \bibinfo {author} {\bibfnamefont {L.}~\bibnamefont {Sansoni}}, \bibinfo
  {author} {\bibfnamefont {S.}~\bibnamefont {Barkhofen}}, \bibinfo {author}
  {\bibfnamefont {C.}~\bibnamefont {Silberhorn}}, \ and\ \bibinfo {author}
  {\bibfnamefont {I.}~\bibnamefont {Jex}},\ }\href@noop {} {\bibfield
  {journal} {\bibinfo  {journal} {Physical Review Letters}\ }\textbf {\bibinfo
  {volume} {119}},\ \bibinfo {pages} {170501} (\bibinfo {year}
  {2017})}\BibitemShut {NoStop}%
\bibitem [{\citenamefont {Scheel}(2004)}]{scheel2004permanents}%
  \BibitemOpen
  \bibfield  {author} {\bibinfo {author} {\bibfnamefont {S.}~\bibnamefont
  {Scheel}},\ }\href@noop {} {\bibfield  {journal} {\bibinfo  {journal} {arXiv
  preprint quant-ph/0406127}\ } (\bibinfo {year} {2004})}\BibitemShut {NoStop}%
\bibitem [{\citenamefont {Caianiello}(1953)}]{Caianiello:1953p12510}%
  \BibitemOpen
  \bibfield  {author} {\bibinfo {author} {\bibfnamefont {E.~R.}\ \bibnamefont
  {Caianiello}},\ }\href@noop {} {\bibfield  {journal} {\bibinfo  {journal} {Il
  Nuovo Cimento}\ }\textbf {\bibinfo {volume} {10}},\ \bibinfo {pages} {1}
  (\bibinfo {year} {1953})}\BibitemShut {NoStop}%
\bibitem [{\citenamefont {Caianiello}(1973)}]{Caianiello73}%
  \BibitemOpen
  \bibfield  {author} {\bibinfo {author} {\bibfnamefont {E.~R.}\ \bibnamefont
  {Caianiello}},\ }\href@noop {} {\emph {\bibinfo {title} {Combinatorics and
  Renormalization in Quantum Field Theory}}}\ (\bibinfo  {publisher} {W. A.
  Benjamin, Inc.},\ \bibinfo {year} {1973})\BibitemShut {NoStop}%
\bibitem [{\citenamefont {Ferraro}\ \emph {et~al.}(2005)\citenamefont
  {Ferraro}, \citenamefont {Olivares},\ and\ \citenamefont
  {Paris}}]{ferraro2005}%
  \BibitemOpen
  \bibfield  {author} {\bibinfo {author} {\bibfnamefont {A.}~\bibnamefont
  {Ferraro}}, \bibinfo {author} {\bibfnamefont {S.}~\bibnamefont {Olivares}}, \
  and\ \bibinfo {author} {\bibfnamefont {M.~G.}\ \bibnamefont {Paris}},\
  }\href@noop {} {\bibfield  {journal} {\bibinfo  {journal} {arXiv preprint
  quant-ph/0503237}\ } (\bibinfo {year} {2005})}\BibitemShut {NoStop}%
\bibitem [{\citenamefont {Barnett}\ and\ \citenamefont
  {Radmore}(1996)}]{Barnett_Radmore}%
  \BibitemOpen
  \bibfield  {author} {\bibinfo {author} {\bibfnamefont {S.~M.}\ \bibnamefont
  {Barnett}}\ and\ \bibinfo {author} {\bibfnamefont {P.}~\bibnamefont
  {Radmore}},\ }\href@noop {} {\emph {\bibinfo {title} {Methods in Theoretical
  Quantum Optics}}}\ (\bibinfo  {publisher} {OUP},\ \bibinfo {year}
  {1996})\BibitemShut {NoStop}%
\bibitem [{\citenamefont {Schleich}(2011)}]{schleich2011quantum}%
  \BibitemOpen
  \bibfield  {author} {\bibinfo {author} {\bibfnamefont {W.~P.}\ \bibnamefont
  {Schleich}},\ }\href@noop {} {\emph {\bibinfo {title} {Quantum optics in
  phase space}}}\ (\bibinfo  {publisher} {John Wiley \& Sons},\ \bibinfo {year}
  {2011})\BibitemShut {NoStop}%
\bibitem [{\citenamefont {Dodonov}\ \emph {et~al.}(1994)\citenamefont
  {Dodonov}, \citenamefont {Man'ko},\ and\ \citenamefont
  {Man'ko}}]{Dodonov:1994p11102}%
  \BibitemOpen
  \bibfield  {author} {\bibinfo {author} {\bibfnamefont {V.~V.}\ \bibnamefont
  {Dodonov}}, \bibinfo {author} {\bibfnamefont {O.~V.}\ \bibnamefont {Man'ko}},
  \ and\ \bibinfo {author} {\bibfnamefont {V.~I.}\ \bibnamefont {Man'ko}},\
  }\href {\doibase 10.1103/PhysRevA.49.2993} {\bibfield  {journal} {\bibinfo
  {journal} {Physical Review A}\ }\textbf {\bibinfo {volume} {49}},\ \bibinfo
  {pages} {2993} (\bibinfo {year} {1994})}\BibitemShut {NoStop}%
\bibitem [{\citenamefont {Husimi}(1940)}]{husimi_formal_1940}%
  \BibitemOpen
  \bibfield  {author} {\bibinfo {author} {\bibfnamefont {K.}~\bibnamefont
  {Husimi}},\ }\href@noop {} {\bibfield  {journal} {\bibinfo  {journal}
  {Proceedings of the Physico-Mathematical Society of Japan. 3rd Series}\
  }\textbf {\bibinfo {volume} {22}},\ \bibinfo {pages} {264} (\bibinfo {year}
  {1940})}\BibitemShut {NoStop}%
\bibitem [{\citenamefont {Glauber}(1963)}]{glauber_coherent_1963}%
  \BibitemOpen
  \bibfield  {author} {\bibinfo {author} {\bibfnamefont {R.~J.}\ \bibnamefont
  {Glauber}},\ }\href {\doibase 10.1103/PhysRev.131.2766} {\bibfield  {journal}
  {\bibinfo  {journal} {Physical Review}\ }\textbf {\bibinfo {volume} {131}},\
  \bibinfo {pages} {2766} (\bibinfo {year} {1963})}\BibitemShut {NoStop}%
\bibitem [{\citenamefont {Sudarshan}(1963)}]{sudarshan_equivalence_1963}%
  \BibitemOpen
  \bibfield  {author} {\bibinfo {author} {\bibfnamefont {E.~C.~G.}\
  \bibnamefont {Sudarshan}},\ }\href {\doibase 10.1103/PhysRevLett.10.277}
  {\bibfield  {journal} {\bibinfo  {journal} {Physical Review Letters}\
  }\textbf {\bibinfo {volume} {10}},\ \bibinfo {pages} {277} (\bibinfo {year}
  {1963})}\BibitemShut {NoStop}%
\bibitem [{\citenamefont {Simon}\ \emph {et~al.}(1994)\citenamefont {Simon},
  \citenamefont {Mukunda},\ and\ \citenamefont {Dutta}}]{Simon:1994p4225}%
  \BibitemOpen
  \bibfield  {author} {\bibinfo {author} {\bibfnamefont {R.}~\bibnamefont
  {Simon}}, \bibinfo {author} {\bibfnamefont {N.}~\bibnamefont {Mukunda}}, \
  and\ \bibinfo {author} {\bibfnamefont {B.}~\bibnamefont {Dutta}},\ }\href
  {\doibase 10.1103/PhysRevA.49.1567} {\bibfield  {journal} {\bibinfo
  {journal} {Physical Review A}\ }\textbf {\bibinfo {volume} {49}},\ \bibinfo
  {pages} {1567} (\bibinfo {year} {1994})}\BibitemShut {NoStop}%
\bibitem [{\citenamefont {Gardiner}\ and\ \citenamefont
  {Zoller}(2004)}]{Gardiner_Zoller}%
  \BibitemOpen
  \bibfield  {author} {\bibinfo {author} {\bibfnamefont {C.}~\bibnamefont
  {Gardiner}}\ and\ \bibinfo {author} {\bibfnamefont {P.}~\bibnamefont
  {Zoller}},\ }\href@noop {} {\emph {\bibinfo {title} {Quantum noise: a
  handbook of Markovian and non-Markovian quantum stochastic methods with
  applications to quantum optics}}},\ Vol.~\bibinfo {volume} {56}\ (\bibinfo
  {publisher} {Springer Science \& Business Media},\ \bibinfo {year}
  {2004})\BibitemShut {NoStop}%
\bibitem [{\citenamefont {Comtet}(1974)}]{Comtet74}%
  \BibitemOpen
  \bibfield  {author} {\bibinfo {author} {\bibfnamefont {L.}~\bibnamefont
  {Comtet}},\ }\href@noop {} {\emph {\bibinfo {title} {Advanced
  Combinatorics}}}\ (\bibinfo  {publisher} {D. Reidel Publishing Company},\
  \bibinfo {year} {1974})\BibitemShut {NoStop}%
\bibitem [{\citenamefont {Hardy}(2006)}]{hardy2006combinatorics}%
  \BibitemOpen
  \bibfield  {author} {\bibinfo {author} {\bibfnamefont {M.}~\bibnamefont
  {Hardy}},\ }\href@noop {} {\bibfield  {journal} {\bibinfo  {journal}
  {Electron. J. Combin}\ }\textbf {\bibinfo {volume} {13}},\ \bibinfo {pages}
  {13} (\bibinfo {year} {2006})}\BibitemShut {NoStop}%
\bibitem [{\citenamefont {Callan}(2009)}]{callan2009combinatorial}%
  \BibitemOpen
  \bibfield  {author} {\bibinfo {author} {\bibfnamefont {D.}~\bibnamefont
  {Callan}},\ }\href@noop {} {\bibfield  {journal} {\bibinfo  {journal} {arXiv
  preprint arXiv:0906.1317}\ } (\bibinfo {year} {2009})}\BibitemShut {NoStop}%
\bibitem [{\citenamefont {Minc}(1978)}]{Minc78}%
  \BibitemOpen
  \bibfield  {author} {\bibinfo {author} {\bibfnamefont {H.}~\bibnamefont
  {Minc}},\ }\href@noop {} {\emph {\bibinfo {title} {Permanents}}}\ (\bibinfo
  {publisher} {Addison-Wesley},\ \bibinfo {year} {1978})\BibitemShut {NoStop}%
\bibitem [{\citenamefont {Valiant}(1979)}]{Valiant:1979p11225}%
  \BibitemOpen
  \bibfield  {author} {\bibinfo {author} {\bibfnamefont {L.}~\bibnamefont
  {Valiant}},\ }\href@noop {} {\bibfield  {journal} {\bibinfo  {journal}
  {Theoretical computer science}\ }\textbf {\bibinfo {volume} {8}},\ \bibinfo
  {pages} {189} (\bibinfo {year} {1979})}\BibitemShut {NoStop}%
\bibitem [{\citenamefont {Chakhmakhchyan}\ \emph {et~al.}(2017)\citenamefont
  {Chakhmakhchyan}, \citenamefont {Cerf},\ and\ \citenamefont
  {Garcia-Patron}}]{chakhmakhchyan_perm_therm}%
  \BibitemOpen
  \bibfield  {author} {\bibinfo {author} {\bibfnamefont {L.}~\bibnamefont
  {Chakhmakhchyan}}, \bibinfo {author} {\bibfnamefont {N.~J.}\ \bibnamefont
  {Cerf}}, \ and\ \bibinfo {author} {\bibfnamefont {R.}~\bibnamefont
  {Garcia-Patron}},\ }\href {\doibase 10.1103/PhysRevA.96.022329} {\bibfield
  {journal} {\bibinfo  {journal} {Physical Review A}\ }\textbf {\bibinfo
  {volume} {96}},\ \bibinfo {pages} {022329} (\bibinfo {year}
  {2017})}\BibitemShut {NoStop}%
\bibitem [{\citenamefont {Kan}(2008)}]{Kan2008}%
  \BibitemOpen
  \bibfield  {author} {\bibinfo {author} {\bibfnamefont {R.}~\bibnamefont
  {Kan}},\ }\href@noop {} {\bibfield  {journal} {\bibinfo  {journal} {Journal
  of Multivariate Analysis}\ }\textbf {\bibinfo {volume} {99}},\ \bibinfo
  {pages} {542} (\bibinfo {year} {2008})}\BibitemShut {NoStop}%
\bibitem [{\citenamefont {Stockmeyer}(1983)}]{stockmeyer_complexity_1983}%
  \BibitemOpen
  \bibfield  {author} {\bibinfo {author} {\bibfnamefont {L.}~\bibnamefont
  {Stockmeyer}},\ }in\ \href {\doibase 10.1145/800061.808740} {\emph {\bibinfo
  {booktitle} {Proceedings of the {Fifteenth} {Annual} {ACM} {Symposium} on
  {Theory} of {Computing}}}},\ \bibinfo {series and number} {{STOC} '83}\
  (\bibinfo  {publisher} {ACM},\ \bibinfo {address} {New York, NY, USA},\
  \bibinfo {year} {1983})\ pp.\ \bibinfo {pages} {118--126}\BibitemShut
  {NoStop}%
\bibitem [{\citenamefont {Horn}\ and\ \citenamefont
  {Johnson}(2013)}]{Horn_Johnson}%
  \BibitemOpen
  \bibfield  {author} {\bibinfo {author} {\bibfnamefont {R.~A.}\ \bibnamefont
  {Horn}}\ and\ \bibinfo {author} {\bibfnamefont {C.~R.}\ \bibnamefont
  {Johnson}},\ }\href@noop {} {\emph {\bibinfo {title} {Matrix Analysis 2nd
  Ed.}}}\ (\bibinfo  {publisher} {Cambridge University Press},\ \bibinfo {year}
  {2013})\BibitemShut {NoStop}%
\bibitem [{\citenamefont {Barvinok}(1996)}]{Barvinok96}%
  \BibitemOpen
  \bibfield  {author} {\bibinfo {author} {\bibfnamefont {A.~I.}\ \bibnamefont
  {Barvinok}},\ }\href {\doibase 10.1287/moor.21.1.65} {\bibfield  {journal}
  {\bibinfo  {journal} {Mathematics of Operations Research}\ }\textbf {\bibinfo
  {volume} {21}},\ \bibinfo {pages} {65} (\bibinfo {year} {1996})}\BibitemShut
  {NoStop}%
\bibitem [{\citenamefont {Hilbe}(2011)}]{Hilbe11}%
  \BibitemOpen
  \bibfield  {author} {\bibinfo {author} {\bibfnamefont {J.~M.}\ \bibnamefont
  {Hilbe}},\ }\href@noop {} {\emph {\bibinfo {title} {Negative Binomial
  Regression}}}\ (\bibinfo  {publisher} {Cambridge University Press},\ \bibinfo
  {year} {2011})\BibitemShut {NoStop}%
\bibitem [{\citenamefont {Bj{\"o}rklund}(2012)}]{bjorklund2012counting}%
  \BibitemOpen
  \bibfield  {author} {\bibinfo {author} {\bibfnamefont {A.}~\bibnamefont
  {Bj{\"o}rklund}},\ }in\ \href@noop {} {\emph {\bibinfo {booktitle}
  {Proceedings of the twenty-third annual ACM-SIAM symposium on Discrete
  Algorithms}}}\ (\bibinfo {organization} {SIAM},\ \bibinfo {year} {2012})\
  pp.\ \bibinfo {pages} {914--921}\BibitemShut {NoStop}%
\bibitem [{\citenamefont {Bentivegna}\ \emph {et~al.}(2015)\citenamefont
  {Bentivegna}, \citenamefont {Spagnolo}, \citenamefont {Vitelli},
  \citenamefont {Flamini}, \citenamefont {Viggianiello}, \citenamefont
  {Latmiral}, \citenamefont {Mataloni}, \citenamefont {Brod}, \citenamefont
  {Galv{\~a}o}, \citenamefont {Crespi} \emph
  {et~al.}}]{bentivegna2015experimental}%
  \BibitemOpen
  \bibfield  {author} {\bibinfo {author} {\bibfnamefont {M.}~\bibnamefont
  {Bentivegna}}, \bibinfo {author} {\bibfnamefont {N.}~\bibnamefont
  {Spagnolo}}, \bibinfo {author} {\bibfnamefont {C.}~\bibnamefont {Vitelli}},
  \bibinfo {author} {\bibfnamefont {F.}~\bibnamefont {Flamini}}, \bibinfo
  {author} {\bibfnamefont {N.}~\bibnamefont {Viggianiello}}, \bibinfo {author}
  {\bibfnamefont {L.}~\bibnamefont {Latmiral}}, \bibinfo {author}
  {\bibfnamefont {P.}~\bibnamefont {Mataloni}}, \bibinfo {author}
  {\bibfnamefont {D.~J.}\ \bibnamefont {Brod}}, \bibinfo {author}
  {\bibfnamefont {E.~F.}\ \bibnamefont {Galv{\~a}o}}, \bibinfo {author}
  {\bibfnamefont {A.}~\bibnamefont {Crespi}},  \emph {et~al.},\ }\href
  {\doibase 10.1126/sciadv.1400255} {\bibfield  {journal} {\bibinfo  {journal}
  {Science Advances}\ }\textbf {\bibinfo {volume} {1}},\ \bibinfo {pages}
  {e1400255} (\bibinfo {year} {2015})}\BibitemShut {NoStop}%
\bibitem [{\citenamefont {Harder}\ \emph {et~al.}(2013)\citenamefont {Harder},
  \citenamefont {Ansari}, \citenamefont {Brecht}, \citenamefont {Dirmeier},
  \citenamefont {Marquardt},\ and\ \citenamefont
  {Silberhorn}}]{harder_optimized_2013}%
  \BibitemOpen
  \bibfield  {author} {\bibinfo {author} {\bibfnamefont {G.}~\bibnamefont
  {Harder}}, \bibinfo {author} {\bibfnamefont {V.}~\bibnamefont {Ansari}},
  \bibinfo {author} {\bibfnamefont {B.}~\bibnamefont {Brecht}}, \bibinfo
  {author} {\bibfnamefont {T.}~\bibnamefont {Dirmeier}}, \bibinfo {author}
  {\bibfnamefont {C.}~\bibnamefont {Marquardt}}, \ and\ \bibinfo {author}
  {\bibfnamefont {C.}~\bibnamefont {Silberhorn}},\ }\href {\doibase
  10.1364/OE.21.013975} {\bibfield  {journal} {\bibinfo  {journal} {Optics
  Express}\ }\textbf {\bibinfo {volume} {21}},\ \bibinfo {pages} {13975}
  (\bibinfo {year} {2013})}\BibitemShut {NoStop}%
\bibitem [{\citenamefont {Harder}\ \emph {et~al.}(2016)\citenamefont {Harder},
  \citenamefont {Bartley}, \citenamefont {Lita}, \citenamefont {Nam},
  \citenamefont {Gerrits},\ and\ \citenamefont
  {Silberhorn}}]{harder_single-mode_2016}%
  \BibitemOpen
  \bibfield  {author} {\bibinfo {author} {\bibfnamefont {G.}~\bibnamefont
  {Harder}}, \bibinfo {author} {\bibfnamefont {T.~J.}\ \bibnamefont {Bartley}},
  \bibinfo {author} {\bibfnamefont {A.~E.}\ \bibnamefont {Lita}}, \bibinfo
  {author} {\bibfnamefont {S.~W.}\ \bibnamefont {Nam}}, \bibinfo {author}
  {\bibfnamefont {T.}~\bibnamefont {Gerrits}}, \ and\ \bibinfo {author}
  {\bibfnamefont {C.}~\bibnamefont {Silberhorn}},\ }\href {\doibase
  10.1103/PhysRevLett.116.143601} {\bibfield  {journal} {\bibinfo  {journal}
  {Physical Review Letters}\ }\textbf {\bibinfo {volume} {116}},\ \bibinfo
  {pages} {143601} (\bibinfo {year} {2016})}\BibitemShut {NoStop}%
\bibitem [{\citenamefont {Rohde}\ \emph {et~al.}(2015)\citenamefont {Rohde},
  \citenamefont {Motes}, \citenamefont {Knott}, \citenamefont {Fitzsimons},
  \citenamefont {Munro},\ and\ \citenamefont {Dowling}}]{Rohde:2015p11547}%
  \BibitemOpen
  \bibfield  {author} {\bibinfo {author} {\bibfnamefont {P.~P.}\ \bibnamefont
  {Rohde}}, \bibinfo {author} {\bibfnamefont {K.~R.}\ \bibnamefont {Motes}},
  \bibinfo {author} {\bibfnamefont {P.~A.}\ \bibnamefont {Knott}}, \bibinfo
  {author} {\bibfnamefont {J.}~\bibnamefont {Fitzsimons}}, \bibinfo {author}
  {\bibfnamefont {W.~J.}\ \bibnamefont {Munro}}, \ and\ \bibinfo {author}
  {\bibfnamefont {J.~P.}\ \bibnamefont {Dowling}},\ }\href {\doibase
  10.1103/PhysRevA.91.012342} {\bibfield  {journal} {\bibinfo  {journal}
  {Physical Review A}\ }\textbf {\bibinfo {volume} {91}},\ \bibinfo {pages}
  {012342} (\bibinfo {year} {2015})}\BibitemShut {NoStop}%
\bibitem [{\citenamefont {Olson}\ \emph {et~al.}(2015)\citenamefont {Olson},
  \citenamefont {Seshadreesan}, \citenamefont {Motes}, \citenamefont {Rohde},\
  and\ \citenamefont {Dowling}}]{Olson:2015p10963}%
  \BibitemOpen
  \bibfield  {author} {\bibinfo {author} {\bibfnamefont {J.~P.}\ \bibnamefont
  {Olson}}, \bibinfo {author} {\bibfnamefont {K.~P.}\ \bibnamefont
  {Seshadreesan}}, \bibinfo {author} {\bibfnamefont {K.~R.}\ \bibnamefont
  {Motes}}, \bibinfo {author} {\bibfnamefont {P.~P.}\ \bibnamefont {Rohde}}, \
  and\ \bibinfo {author} {\bibfnamefont {J.~P.}\ \bibnamefont {Dowling}},\
  }\href {\doibase 10.1103/PhysRevA.91.022317} {\bibfield  {journal} {\bibinfo
  {journal} {Physical Review A}\ }\textbf {\bibinfo {volume} {91}},\ \bibinfo
  {pages} {022317} (\bibinfo {year} {2015})}\BibitemShut {NoStop}%
\bibitem [{\citenamefont {Seshadreesan}\ \emph {et~al.}(2015)\citenamefont
  {Seshadreesan}, \citenamefont {Olson}, \citenamefont {Motes}, \citenamefont
  {Rohde},\ and\ \citenamefont {Dowling}}]{Seshadreesan:2015p12695}%
  \BibitemOpen
  \bibfield  {author} {\bibinfo {author} {\bibfnamefont {K.~P.}\ \bibnamefont
  {Seshadreesan}}, \bibinfo {author} {\bibfnamefont {J.~P.}\ \bibnamefont
  {Olson}}, \bibinfo {author} {\bibfnamefont {K.~R.}\ \bibnamefont {Motes}},
  \bibinfo {author} {\bibfnamefont {P.~P.}\ \bibnamefont {Rohde}}, \ and\
  \bibinfo {author} {\bibfnamefont {J.~P.}\ \bibnamefont {Dowling}},\ }\href
  {\doibase 10.1103/PhysRevA.91.022334} {\bibfield  {journal} {\bibinfo
  {journal} {Physical Review A}\ }\textbf {\bibinfo {volume} {91}},\ \bibinfo
  {pages} {022334} (\bibinfo {year} {2015})}\BibitemShut {NoStop}%
\bibitem [{\citenamefont {Chakhmakhchyan}\ and\ \citenamefont
  {Cerf}(2017)}]{chakhmakhchyan2017}%
  \BibitemOpen
  \bibfield  {author} {\bibinfo {author} {\bibfnamefont {L.}~\bibnamefont
  {Chakhmakhchyan}}\ and\ \bibinfo {author} {\bibfnamefont {N.~J.}\
  \bibnamefont {Cerf}},\ }\href {\doibase 10.1103/PhysRevA.96.032326}
  {\bibfield  {journal} {\bibinfo  {journal} {Physical Review A}\ }\textbf
  {\bibinfo {volume} {96}},\ \bibinfo {pages} {032326} (\bibinfo {year}
  {2017})}\BibitemShut {NoStop}%
\bibitem [{\citenamefont {Lund}\ \emph {et~al.}(2017)\citenamefont {Lund},
  \citenamefont {Rahimi-Keshari},\ and\ \citenamefont
  {Ralph}}]{lund_exact_2017}%
  \BibitemOpen
  \bibfield  {author} {\bibinfo {author} {\bibfnamefont {A.~P.}\ \bibnamefont
  {Lund}}, \bibinfo {author} {\bibfnamefont {S.}~\bibnamefont
  {Rahimi-Keshari}}, \ and\ \bibinfo {author} {\bibfnamefont {T.~C.}\
  \bibnamefont {Ralph}},\ }\href {\doibase 10.1103/PhysRevA.96.022301}
  {\bibfield  {journal} {\bibinfo  {journal} {Physical Review A}\ }\textbf
  {\bibinfo {volume} {96}},\ \bibinfo {pages} {022301} (\bibinfo {year}
  {2017})}\BibitemShut {NoStop}%
\bibitem [{\citenamefont {Chabaud}\ \emph {et~al.}(2017)\citenamefont
  {Chabaud}, \citenamefont {Douce}, \citenamefont {Markham}, \citenamefont {van
  Loock}, \citenamefont {Kashefi},\ and\ \citenamefont
  {Ferrini}}]{chabaud2017}%
  \BibitemOpen
  \bibfield  {author} {\bibinfo {author} {\bibfnamefont {U.}~\bibnamefont
  {Chabaud}}, \bibinfo {author} {\bibfnamefont {T.}~\bibnamefont {Douce}},
  \bibinfo {author} {\bibfnamefont {D.}~\bibnamefont {Markham}}, \bibinfo
  {author} {\bibfnamefont {P.}~\bibnamefont {van Loock}}, \bibinfo {author}
  {\bibfnamefont {E.}~\bibnamefont {Kashefi}}, \ and\ \bibinfo {author}
  {\bibfnamefont {G.}~\bibnamefont {Ferrini}},\ }\href {\doibase
  10.1103/PhysRevA.96.062307} {\bibfield  {journal} {\bibinfo  {journal}
  {Physical Review A}\ }\textbf {\bibinfo {volume} {96}},\ \bibinfo {pages}
  {062307} (\bibinfo {year} {2017})}\BibitemShut {NoStop}%
\bibitem [{\citenamefont {Gogolin}\ \emph {et~al.}(2013)\citenamefont
  {Gogolin}, \citenamefont {Kliesch}, \citenamefont {Aolita},\ and\
  \citenamefont {Eisert}}]{gogolin2013boson}%
  \BibitemOpen
  \bibfield  {author} {\bibinfo {author} {\bibfnamefont {C.}~\bibnamefont
  {Gogolin}}, \bibinfo {author} {\bibfnamefont {M.}~\bibnamefont {Kliesch}},
  \bibinfo {author} {\bibfnamefont {L.}~\bibnamefont {Aolita}}, \ and\ \bibinfo
  {author} {\bibfnamefont {J.}~\bibnamefont {Eisert}},\ }\href@noop {}
  {\bibfield  {journal} {\bibinfo  {journal} {arXiv preprint arXiv:1306.3995}\
  } (\bibinfo {year} {2013})}\BibitemShut {NoStop}%
\bibitem [{\citenamefont {Aaronson}\ and\ \citenamefont
  {Arkhipov}(2013{\natexlab{b}})}]{aaronson2013bosonsampling}%
  \BibitemOpen
  \bibfield  {author} {\bibinfo {author} {\bibfnamefont {S.}~\bibnamefont
  {Aaronson}}\ and\ \bibinfo {author} {\bibfnamefont {A.}~\bibnamefont
  {Arkhipov}},\ }\href@noop {} {\bibfield  {journal} {\bibinfo  {journal}
  {arXiv preprint arXiv:1309.7460}\ } (\bibinfo {year}
  {2013}{\natexlab{b}})}\BibitemShut {NoStop}%
\bibitem [{\citenamefont {Carolan}\ \emph {et~al.}(2014)\citenamefont
  {Carolan}, \citenamefont {Meinecke}, \citenamefont {Shadbolt}, \citenamefont
  {Russell}, \citenamefont {Ismail}, \citenamefont {W{\"o}rhoff}, \citenamefont
  {Rudolph}, \citenamefont {Thompson}, \citenamefont {O'Brien}, \citenamefont
  {Matthews},\ and\ \citenamefont {Laing}}]{Carolan2014}%
  \BibitemOpen
  \bibfield  {author} {\bibinfo {author} {\bibfnamefont {J.}~\bibnamefont
  {Carolan}}, \bibinfo {author} {\bibfnamefont {J.~D.~A.}\ \bibnamefont
  {Meinecke}}, \bibinfo {author} {\bibfnamefont {P.~J.}\ \bibnamefont
  {Shadbolt}}, \bibinfo {author} {\bibfnamefont {N.~J.}\ \bibnamefont
  {Russell}}, \bibinfo {author} {\bibfnamefont {N.}~\bibnamefont {Ismail}},
  \bibinfo {author} {\bibfnamefont {K.}~\bibnamefont {W{\"o}rhoff}}, \bibinfo
  {author} {\bibfnamefont {T.}~\bibnamefont {Rudolph}}, \bibinfo {author}
  {\bibfnamefont {M.~G.}\ \bibnamefont {Thompson}}, \bibinfo {author}
  {\bibfnamefont {J.~L.}\ \bibnamefont {O'Brien}}, \bibinfo {author}
  {\bibfnamefont {J.~C.~F.}\ \bibnamefont {Matthews}}, \ and\ \bibinfo {author}
  {\bibfnamefont {A.}~\bibnamefont {Laing}},\ }\href {\doibase
  10.1038/nphoton.2014.152} {\bibfield  {journal} {\bibinfo  {journal} {Nature
  Photonics}\ }\textbf {\bibinfo {volume} {8}},\ \bibinfo {pages} {621}
  (\bibinfo {year} {2014})}\BibitemShut {NoStop}%
\bibitem [{\citenamefont {Spagnolo}\ \emph {et~al.}(2014)\citenamefont
  {Spagnolo}, \citenamefont {Vitelli}, \citenamefont {Bentivegna},
  \citenamefont {Brod}, \citenamefont {Crespi}, \citenamefont {Flamini},
  \citenamefont {Giacomini}, \citenamefont {Milani}, \citenamefont {Ramponi},
  \citenamefont {Mataloni}, \citenamefont {Osellame}, \citenamefont
  {Galv{\~a}o},\ and\ \citenamefont {Sciarrino}}]{Spagnolo:2014p10480}%
  \BibitemOpen
  \bibfield  {author} {\bibinfo {author} {\bibfnamefont {N.}~\bibnamefont
  {Spagnolo}}, \bibinfo {author} {\bibfnamefont {C.}~\bibnamefont {Vitelli}},
  \bibinfo {author} {\bibfnamefont {M.}~\bibnamefont {Bentivegna}}, \bibinfo
  {author} {\bibfnamefont {D.~J.}\ \bibnamefont {Brod}}, \bibinfo {author}
  {\bibfnamefont {A.}~\bibnamefont {Crespi}}, \bibinfo {author} {\bibfnamefont
  {F.}~\bibnamefont {Flamini}}, \bibinfo {author} {\bibfnamefont
  {S.}~\bibnamefont {Giacomini}}, \bibinfo {author} {\bibfnamefont
  {G.}~\bibnamefont {Milani}}, \bibinfo {author} {\bibfnamefont
  {R.}~\bibnamefont {Ramponi}}, \bibinfo {author} {\bibfnamefont
  {P.}~\bibnamefont {Mataloni}}, \bibinfo {author} {\bibfnamefont
  {R.}~\bibnamefont {Osellame}}, \bibinfo {author} {\bibfnamefont {E.~F.}\
  \bibnamefont {Galv{\~a}o}}, \ and\ \bibinfo {author} {\bibfnamefont
  {F.}~\bibnamefont {Sciarrino}},\ }\href {\doibase 10.1038/nphoton.2014.135}
  {\bibfield  {journal} {\bibinfo  {journal} {Nature Photonics}\ }\textbf
  {\bibinfo {volume} {8}},\ \bibinfo {pages} {615} (\bibinfo {year}
  {2014})}\BibitemShut {NoStop}%
\bibitem [{\citenamefont {Bentivegna}\ \emph {et~al.}(2014)\citenamefont
  {Bentivegna}, \citenamefont {Spagnolo}, \citenamefont {Vitelli},
  \citenamefont {Brod}, \citenamefont {Crespi}, \citenamefont {Flamini},
  \citenamefont {Ramponi}, \citenamefont {Mataloni}, \citenamefont {Osellame},
  \citenamefont {Galv{\~a}o} \emph {et~al.}}]{bentivegna2014bayesian}%
  \BibitemOpen
  \bibfield  {author} {\bibinfo {author} {\bibfnamefont {M.}~\bibnamefont
  {Bentivegna}}, \bibinfo {author} {\bibfnamefont {N.}~\bibnamefont
  {Spagnolo}}, \bibinfo {author} {\bibfnamefont {C.}~\bibnamefont {Vitelli}},
  \bibinfo {author} {\bibfnamefont {D.~J.}\ \bibnamefont {Brod}}, \bibinfo
  {author} {\bibfnamefont {A.}~\bibnamefont {Crespi}}, \bibinfo {author}
  {\bibfnamefont {F.}~\bibnamefont {Flamini}}, \bibinfo {author} {\bibfnamefont
  {R.}~\bibnamefont {Ramponi}}, \bibinfo {author} {\bibfnamefont
  {P.}~\bibnamefont {Mataloni}}, \bibinfo {author} {\bibfnamefont
  {R.}~\bibnamefont {Osellame}}, \bibinfo {author} {\bibfnamefont {E.~F.}\
  \bibnamefont {Galv{\~a}o}},  \emph {et~al.},\ }\href@noop {} {\bibfield
  {journal} {\bibinfo  {journal} {International Journal of Quantum
  Information}\ }\textbf {\bibinfo {volume} {12}},\ \bibinfo {pages} {1560028}
  (\bibinfo {year} {2014})}\BibitemShut {NoStop}%
\bibitem [{\citenamefont {Tichy}\ \emph {et~al.}(2014)\citenamefont {Tichy},
  \citenamefont {Mayer}, \citenamefont {Buchleitner},\ and\ \citenamefont
  {M{\o}lmer}}]{Tichy2014}%
  \BibitemOpen
  \bibfield  {author} {\bibinfo {author} {\bibfnamefont {M.~C.}\ \bibnamefont
  {Tichy}}, \bibinfo {author} {\bibfnamefont {K.}~\bibnamefont {Mayer}},
  \bibinfo {author} {\bibfnamefont {A.}~\bibnamefont {Buchleitner}}, \ and\
  \bibinfo {author} {\bibfnamefont {K.}~\bibnamefont {M{\o}lmer}},\ }\href
  {\doibase 10.1103/PhysRevLett.113.020502} {\bibfield  {journal} {\bibinfo
  {journal} {Physical Review Letters}\ }\textbf {\bibinfo {volume} {113}},\
  \bibinfo {pages} {020502} (\bibinfo {year} {2014})}\BibitemShut {NoStop}%
\bibitem [{\citenamefont {Walschaers}\ \emph {et~al.}(2016)\citenamefont
  {Walschaers}, \citenamefont {Kuipers}, \citenamefont {Urbina}, \citenamefont
  {Mayer}, \citenamefont {Tichy}, \citenamefont {Richter},\ and\ \citenamefont
  {Buchleitner}}]{walschaers2016statistical}%
  \BibitemOpen
  \bibfield  {author} {\bibinfo {author} {\bibfnamefont {M.}~\bibnamefont
  {Walschaers}}, \bibinfo {author} {\bibfnamefont {J.}~\bibnamefont {Kuipers}},
  \bibinfo {author} {\bibfnamefont {J.-D.}\ \bibnamefont {Urbina}}, \bibinfo
  {author} {\bibfnamefont {K.}~\bibnamefont {Mayer}}, \bibinfo {author}
  {\bibfnamefont {M.~C.}\ \bibnamefont {Tichy}}, \bibinfo {author}
  {\bibfnamefont {K.}~\bibnamefont {Richter}}, \ and\ \bibinfo {author}
  {\bibfnamefont {A.}~\bibnamefont {Buchleitner}},\ }\href {\doibase
  10.1088/1367-2630/18/3/032001} {\bibfield  {journal} {\bibinfo  {journal}
  {New Journal of Physics}\ }\textbf {\bibinfo {volume} {18}},\ \bibinfo
  {pages} {032001} (\bibinfo {year} {2016})}\BibitemShut {NoStop}%
\bibitem [{\citenamefont {\ifmmode \check{R}\else
  \v{R}\fi{}eh\'a\ifmmode~\check{c}\else \v{c}\fi{}ek}\ \emph
  {et~al.}(2009)\citenamefont {\ifmmode \check{R}\else
  \v{R}\fi{}eh\'a\ifmmode~\check{c}\else \v{c}\fi{}ek}, \citenamefont
  {Olivares}, \citenamefont {Mogilevtsev}, \citenamefont {Hradil},
  \citenamefont {Paris}, \citenamefont {Fornaro}, \citenamefont {D'Auria},
  \citenamefont {Porzio},\ and\ \citenamefont {Solimeno}}]{Rehacek2009}%
  \BibitemOpen
  \bibfield  {author} {\bibinfo {author} {\bibfnamefont {J.}~\bibnamefont
  {\ifmmode \check{R}\else \v{R}\fi{}eh\'a\ifmmode~\check{c}\else
  \v{c}\fi{}ek}}, \bibinfo {author} {\bibfnamefont {S.}~\bibnamefont
  {Olivares}}, \bibinfo {author} {\bibfnamefont {D.}~\bibnamefont
  {Mogilevtsev}}, \bibinfo {author} {\bibfnamefont {Z.}~\bibnamefont {Hradil}},
  \bibinfo {author} {\bibfnamefont {M.~G.~A.}\ \bibnamefont {Paris}}, \bibinfo
  {author} {\bibfnamefont {S.}~\bibnamefont {Fornaro}}, \bibinfo {author}
  {\bibfnamefont {V.}~\bibnamefont {D'Auria}}, \bibinfo {author} {\bibfnamefont
  {A.}~\bibnamefont {Porzio}}, \ and\ \bibinfo {author} {\bibfnamefont
  {S.}~\bibnamefont {Solimeno}},\ }\href {\doibase 10.1103/PhysRevA.79.032111}
  {\bibfield  {journal} {\bibinfo  {journal} {Physical Review A}\ }\textbf
  {\bibinfo {volume} {79}},\ \bibinfo {pages} {032111} (\bibinfo {year}
  {2009})}\BibitemShut {NoStop}%
\end{thebibliography}%


\end{document}